\documentclass[pra, twocolumn, superscriptaddress]{revtex4-2}
\usepackage[dvipdfmx]{graphicx}
\usepackage[utf8]{inputenc}
\usepackage{amsmath}
\usepackage{amssymb}
\usepackage{bm}
\usepackage{bbm}
\usepackage{mathtools}
\usepackage{braket}
\usepackage{qcircuit}
\usepackage{url}
\usepackage{hyperref}
\usepackage{color} \usepackage[subrefformat=parens]{subcaption} 
\hypersetup{
    colorlinks,
    citecolor=blue,
    linkcolor=blue,
    urlcolor=blue
}

\xyoption{color}

\newcommand{\floor}[1]{\left\lfloor#1\right\rfloor}
\newcommand{\ceil}[1]{\left\lceil#1\right\rceil}

\newcommand{\ifdebug}[1]{#1}
\renewcommand{\ifdebug}[1]{} 

\ifdebug{
\usepackage{color}
\usepackage{soul}
\usepackage{ulem}
\usepackage[whole]{bxcjkjatype}
\definecolor{mycyan}{RGB}{0,8,255}
\definecolor{mypink}{cmyk}{0,0.7808,0.4429,0.1412}
\definecolor{mygreen}{RGB}{0,102,51}
\definecolor{myorange}{RGB}{255,160,0}

}

\begin{document}

\title{ISAAQ: Ising Machine Assisted Quantum Compiler}

\author{Soshun Naito}
\affiliation{
Department of Information and Communication Engineering, Graduate School of Information Science and Technology, The University of Tokyo, 7-3-1 Hongo, Bunkyo-ku, Tokyo 113-8656, Japan
}

\author{Yoshihiko Hasegawa}
\affiliation{
Department of Information and Communication Engineering, Graduate School of Information Science and Technology, The University of Tokyo, 7-3-1 Hongo, Bunkyo-ku, Tokyo 113-8656, Japan
}

\author{Yoshiki Matsuda}
\affiliation{
Fixstars Corporation, 3-1-1 Shibaura, Minato-ku, Tokyo 108-0023, Japan
}
\affiliation{
Green Computing System Research Organization, Waseda University, 27 Wasedacho, Shinjuku-ku, Tokyo 162-0042, Japan
}

\author{Shu Tanaka}
\affiliation{
Green Computing System Research Organization, Waseda University, 27 Wasedacho, Shinjuku-ku, Tokyo 162-0042, Japan
}
\affiliation{
Department of Applied Physics and Physico-Informatics, Keio University, 3-14-1 Hiyoshi, Kohoku-ku, Yokohama, Kanagawa
223-8522, Japan
}

\begin{abstract}

It is imperative to compile quantum circuits for Noisy Intermediate-Scale Quantum (NISQ) devices because of the limited connectivity of physical qubits and the high error rates of gate operations.
One of the most critical steps in quantum circuit compilation is qubit routing, an NP-Hard problem that involves placing and moving logical qubits to minimize compilation overhead.
In this study, we propose ISing mAchine Assisted Quantum compiler (ISAAQ) to perform qubit routing with Ising machines, which can efficiently solve Quadratic Unconstrained Binary Optimization (QUBO) problems.
ISAAQ accurately estimates the compilation costs by updating itself using previous compilation results, and accelerates qubit routing by solving QUBO problems in parallel with multiple Ising machines.
In addition, ISAAQ exploits a cost-reduction method that implements commutative logical Controlled-NOT (CNOT) gates with fewer physical CNOT gates, which is particularly effective for planar devices when implementing original gates.
Experimental results on both IBM QX5 and IBM QX20 show that ISAAQ outperforms the heuristic methods available in Qiskit and tket, as well as an existing QUBO method, requiring fewer physical CNOT gates for most benchmark circuits.
ISAAQ performs particularly well on large circuits, demonstrating its strong scalability with respect to the number of logical CNOT gates.

\end{abstract} 
\maketitle

\section{Introduction}
Quantum computation is a rapidly growing technology that uses quantum properties to perform computations efficiently.
Quantum algorithms outperform classical algorithms for certain tasks, including prime factorization~\cite{shor1999polynomial} and database search~\cite{grover1996fast}.
Variational Quantum Eigensolver (VQE) and Quantum Approximate Optimization Algorithm (QAOA) are effective tools for solving optimization problems in a variety of areas, including chemistry~\cite{cao2019quantum} and machine learning~\cite{biamonte2017quantum}.

Over the past decade, the infrastructure for quantum computation has been established and is expected to continue to grow rapidly.
Currently, the most widely used quantum devices are called Noisy Intermediate-Scale Quantum (NISQ) devices, which contain between tens and hundreds of physical qubits.
These devices can be accessed through cloud services such as IBM Quantum~\cite{ibm-quantum} and Amazon Bracket~\cite{amazon-bracket}, allowing researchers and developers to run quantum circuits without the need to procure their own hardware.

Despite the progress made in recent years, NISQ devices remain in the early stages of development and face numerous challenges.
For instance, the quantum gates that NISQ devices can execute directly are limited to single-input gates and specific multiple-input gates such as controlled-NOT (CNOT) gates.
In addition, hardware-supported multiple-input gates can only be applied to adjacent physical qubits, restricting the types of quantum circuits that can be efficiently executed on these devices.
Furthermore, NISQ devices are prone to errors, which can accumulate over the course of a computation and degrade overall fidelity.

To perform quantum computation accurately, the circuit must be adapted to meet the constraints of the physical device while minimizing errors through a process called quantum circuit compilation.
The process of quantum circuit compilation consists of three steps: gate decomposition, qubit routing, and circuit synthesis.
In gate decomposition, the gates of the original circuit are decomposed into gates that can be executed directly on the hardware.
During qubit routing, logical qubits are placed on physical qubits and moved to minimize compilation overhead.
Circuit synthesis involves creating a logically equivalent circuit using only physical gates.

In this study, we focus on qubit routing.
The overhead associated with qubit routing, which is the number of additional physical CNOT gates needed, comes from the use of remote CNOT gates and SWAP gates.
Remote CNOT gates allow CNOT gates to be applied to distant physical qubits, however the increase in the distance between qubits increases the number of physical CNOT gates required.
SWAP gates swap the states of adjacent physical qubits, which can be used to move logical qubits virtually.

Previous studies have proposed several approaches to qubit routing, including exact solutions such as dynamic programming~\cite{siraichi2018qubit}, exhaustive search~\cite{zhu2020exact,burgholzer2021limiting}, and solvers such as SAT~\cite{lye2015determining,wille2014optimal}, SMT~\cite{tan2021optimal,wille2019mapping,murali2019noise}, and integer programming~\cite{shafaei2014qubit,de2019finding,nannicini2021optimal}.
However, these exact solutions are only applicable to very small problems due to their exponential computational complexity.
The qubit routing problem cannot be solved in polynomial time for general graphs, making it difficult to solve exactly even in the case of intermediate-scale problems~\cite{botea2018complexity}.

To address this challenge, previous studies have proposed approximate solutions that use methods like A* search~\cite{zhang2020depth,zulehner2018efficient,tannu2019not}, Monte Carlo tree search~\cite{zhou2020monte,sinha2021qubit}, simulated annealing~\cite{niu2020hardware,zhou2020quantum,finigan2018qubit}, reinforcement learning~\cite{fosel2021quantum,pozzi2020using}, and heuristics~\cite{zhu2020dynamic,murali2019noise,cowtan2019qubit,itoko2020optimization,siraichi2018qubit,li2020qubits,ash2019qure,nishio2020extracting,zhu2021iterated,ren2021nuwa,paler2019influence,li2020qubit,li2019tackling,wille2016look}.
Some of these approximate solutions have been implemented in qiskit~\cite{qiskit} and tket~\cite{tket}, and have good scalability with respect to the problem size (i.e., the number of qubits and gates).
The SABRE algorithm, implemented in qiskit, minimizes the number of SWAP gates while updating the placement of logical qubits bidirectionally~\cite{li2019tackling}.
The method proposed by Cowtan et al., which is used in tket, is a heuristic that inserts SWAP gates while checking all CNOT gates from left to right~\cite{cowtan2019qubit}.

Recently, \textit{Ising machines}, efficient solvers specialized in Quadratic Unconstrained Binary Optimization (QUBO) problems have become available~\cite{fixstars, fujitsu, hitachi, d-wave, honjo2021100}.
Ising machines include quantum annealing machines, which are non-Neumann-type computers that physically exploit quantum effects to accelerate computation~\cite{d-wave}.
The key feature of Ising machines is their ability to efficiently find near-optimal solutions to large QUBO problems.
Therefore, Ising machines are expected to be applied to practical, large-scale combinatorial optimization problems.
QUBO problems are defined as follows:
\begin{align}
\mathrm{minimize} \quad &\sum_{i, j} Q_{ij} x_i x_j \nonumber \\
\mathrm{subject \; to} \quad &x_i \in \left\{ 0, 1 \right\},
\end{align}
where $Q_{ij} \in \mathbb{R}$ are the coefficients that represent the problem to be solved.
Due to their rich expressiveness, QUBO problems can represent many combinatorial optimization problems, including NP-Complete and NP-Hard problems such as graph partitioning, vertex cover, 3SAT, job sequencing, TSP, and MAXCUT~\cite{tanahashi2019application,tanaka-book,lucas2014ising}.

Despite their substantial potential for solving combinatorial optimization problems, the application of Ising machines to quantum circuit compilation is still being explored.
For example, Dury et al. formulated the problem of determining the initial placement of logical qubits (called qubit allocation) as a QUBO problem, using heuristically determined coefficients based on error rates~\cite{dury2020qubo}.
Butko et al. followed a different approach, formulating quantum circuit compilation as a task scheduling problem that considered the interactions between logical qubits as tasks~\cite{butko2020tiger}.
However, these methods could not fully exploit the potential of Ising machines.
The QUBO formulation method proposed by Dury et al. has substantial scalability but did not consider the routes of logical qubits, only optimizing the initial placement of the qubits.
The method proposed by Butko et al. has rich expressiveness, but the application of the method was limited to very small problems due to its poor scalability with respect to the number of qubits and gates.

In this study, we introduce \textbf{IS}ing m\textbf{A}chine \textbf{A}ssisted \textbf{Q}uantum compiler (ISAAQ), a novel quantum compiler that overcomes the limitations of existing methods and fully leverages the advantage of Ising machines.
ISAAQ slices a circuit into layers with a configurable upper limit on the number of CNOT gates, allowing it to represent the routes of logical qubits across the circuit while avoiding excessive use of binary variables.
ISAAQ generates a QUBO model based on the number of physical CNOT gates required to decompose remote CNOT and SWAP gates. Because the calculation of the exact number of SWAP gates is computationally complex, ISAAQ uses a QUBO model with optimized parameters to approximate the cost of SWAP gates.
Unlike previous QUBO methods that have used fixed parameters, ISAAQ updates the coefficients of its QUBO model using previous compilation results, improving the accuracy of its approximations over time.

In addition, ISAAQ uses several methods to make quantum circuit compilation more efficient.
ISAAQ converts the qubit routing problem into smaller, more manageable QUBO problems by partitioning the circuit into multiple circuit chunks.
These QUBO problems are adjusted to be small enough for the capabilities of Ising machines.
To speed up qubit routing, ISAAQ can also solve QUBO problems with multiple Ising machines while maintaining a high level of compilation quality.
Finally, ISAAQ employs a method that uses relay qubits to cache computational results when implementing a set of commutative logical CNOT gates with a shared control or target qubit.
This method can minimize the number of physical CNOT gates by removing duplicates.

We evaluate ISAAQ using various benchmark circuits on IBM QX5 and IBM QX20 and compare its performance with that of the heuristic methods (Li et al.~\cite{li2019tackling}, Cowtan et al.~\cite{cowtan2019qubit}) and the QUBO method (Dury et al.~\cite{dury2020qubo}).
Our experimental results show that ISAAQ possesses several unique features including its self-updating capability, parallelizability, and cost-saving ability.
In addition, ISAAQ demonstrates superior compilation performance and is particularly effective for larger circuits, verifying its scalability to the number of logical CNOT gates.

Our main contributions are summarized as follows:
\begin{itemize}
    \item We develop a method for formulating the number of physical CNOT gates required for qubit routing as a QUBO problem, which can be solved using Ising machines. The QUBO model represents the routes of logical qubits along the entire circuit.
    \item We create a method for adaptively updating the QUBO model using previous compilation results. It allows the QUBO model to accurately represent the compilation cost, which is not simple to estimate. We also develop a method for determining the initial coefficients for the QUBO model.
    \item We develop a method for decomposing a large routing problem into smaller problems by partitioning the logical circuit. This feature allows ISAAQ to solve the QUBO problems in parallel using multiple Ising machines while maintaining high compilation quality.
    \item We develop a method for efficiently synthesizing a physical circuit using the routing result. ISAAQ reduces the number of physical CNOT gates when synthesizing commutative logical CNOT gates with a shared control or target qubit.
    \item We implement ISAAQ using Fixstars Amplify Annealing Engine~\cite{fixstars} and demonstrate its advantage over existing methods, for various benchmark circuits and for both IBM QX5 and IBM QX20. The superior performance of ISAAQ is more significant for larger circuits, indicating its ability to optimize qubit routing globally.
\end{itemize}

\section{Background}
\label{section:background}

\subsection{Quantum Computing}
In quantum computing, qubits are the fundamental units of information.
Each qubit can exist in a superposition of two states, referred to as $\ket{0}$ and $\ket{1}$.
In a quantum system containing $N$ qubits, the state vector of the system is a superposition of $2^N$ bases, expressed mathematically as
\begin{align}
\ket{\psi} = \sum_{i_0 \in \{0,1\}} \cdots \sum_{i_{N-1} \in \{0,1\}}
\alpha_{i_0 \cdots i_{N-1}} \ket{i_0 \cdots i_{N-1}},
\end{align}
where the coefficients $\alpha_{i_0 \cdots i_{N-1}}$ represent the amplitudes of the bases.
As the probability of observing each basis $\ket{i_0 \cdots i_{N-1}}$ is equal to $\|\alpha_{i_0 \cdots i_{N-1}}\|^2$, the coefficients satisfy the following condition:
\begin{align}
\sum_{i_0 \in \{0,1\}} \cdots \sum_{i_{N-1} \in \{0,1\}}
\|\alpha_{i_0 \cdots i_{N-1}}\|^2 = 1.
\end{align}

To perform quantum computation, we apply a unitary transformation $U$ to the initial state $\ket{\psi_0}$ to produce the desired final state $\ket{\psi} = U \ket{\psi_0}$.
In cases where $U$ is too complex to be executed directly, it is implemented as a composite of fundamental unitary transformations.
After preparing the final state, the result of the computation can be obtained as a binary string by applying a projection measurement.
The measurement result is not deterministic, since the amplitudes of the respective bases determine the probability distribution.

\begin{figure}[t]
    \begin{subfigure}[]{0.45\linewidth}
        \centering
            \[ \Qcircuit @C=1em @R=1em {
            \lstick{\ket{0}} & \gate{H} & \rstick{\frac{\ket{0}+\ket{1}}{\sqrt{2}}} \qw \\
            \lstick{\ket{1}} & \gate{H} & \rstick{\frac{\ket{0}-\ket{1}}{\sqrt{2}}} \qw
            } \]
        \subcaption{}
        \label{fig:H_gate}
    \end{subfigure}
    \begin{subfigure}[]{0.5\linewidth}
        \centering
            \[ \Qcircuit @C=1em @R=1em {
            \lstick{\ket{0}} & \gate{T} & \rstick{\ket{0}} \qw \\
            \lstick{\ket{1}} & \gate{T} & \rstick{e^{\frac{i\pi}{4}}\ket{1}} \qw
            } \]
        \subcaption{}
        \label{fig:T_gate}
    \end{subfigure}
    
    \begin{subfigure}[]{0.45\linewidth}
        \centering
        \vspace{12pt}
            \[ \Qcircuit @C=1em @R=1em {
            \lstick{\ket{x}} & \ctrl{1} & \rstick{\ket{x}} \qw \\
            \lstick{\ket{y}} & \targ & \rstick{\ket{x \oplus y}} \qw
            } \]
        \captionsetup{width=.9\linewidth}
        \subcaption{}
        \label{fig:CNOT_gate}
    \end{subfigure}
    \begin{subfigure}[]{0.5\linewidth}
        \centering
            \[ \Qcircuit @C=1em @R=1em {
            \lstick{\ket{x}} & \ctrl{1} & \rstick{\ket{x}} \qw \\
            \lstick{\ket{y}} & \ctrl{1} & \rstick{\ket{y}} \qw \\
            \lstick{\ket{z}} & \targ & \rstick{\ket{xy \oplus z}} \qw
            } \]
        \captionsetup{width=.9\linewidth}
        \subcaption{}
        \label{fig:Toffoli_gate}
    \end{subfigure}
    \caption{
        Examples of quantum gates. (a) Hadamard gate. (b) T gate. (c) CNOT gate (flips the target when $x=1$). (d) Toffoli gate (flips the target when $x=y=1$).
    }
    \label{fig:gate_example}
\end{figure}
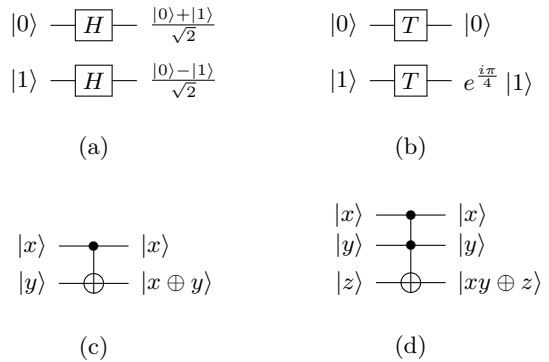

Quantum operations can be visually represented using quantum circuits.
On these circuits, qubits are illustrated as wires and unitary transformations are represented as quantum gates acting on the qubits.
There are two types of quantum gates: single-input gates, such as Pauli gates, Clifford gates, and rotation gates, and multiple-input gates, including CNOT gates and Toffoli gates.
Examples of various quantum gates are depicted in Fig.~\ref{fig:gate_example}.

\subsection{Quantum Circuit Compilation}

\subsubsection{Motivation}
To run a quantum circuit on NISQ devices, the circuit must be compiled specifically for these devices.
Quantum circuits are usually designed for ideal devices that have an unlimited number of fully-connected physical qubits and can apply noiseless unitary gates.

However, NISQ devices have several limitations that must be taken into account.
One of these limitations is the limited set of quantum gates that can be executed directly, which includes single-input gates and a few specific two-input gates.
These two-input gates can only be applied to physically adjacent qubits because they are typically only coupled to their nearest neighbors.
For example, the Sycamore processor designed by Google has 54 qubits arranged in a rectangular lattice and supports controlled-Z (CZ) gates and $i\sqrt{\mathrm{SWAP}}$ gates as two-input gates~\cite{arute2019quantum}.
The processors designed by IBM have qubits placed on a heavy hex lattice and support CNOT gates~\cite{ibm-quantum}.
A further limitation applicable to NISQ devices is that they are prone to errors when applying quantum gates, which accumulate over the course of a computation.
The use of a large number of two-input gates degrades the quality of the computation because of the inaccuracies of two-input gates, compared with single-input gates.

To address these problems, the quantum gates in a circuit must be implemented using only hardware-supported gates while minimizing the number of two-input gates as possible.
This can be achieved by moving interacting logical qubits to render them physically adjacent, or decomposing the original gates into multiple local two-input gates.
Reducing the number of noisy gates may also improve the effectiveness of a computation.

\subsubsection{Gate Decomposition}
Gate decomposition is the process of breaking down large, complex gates into smaller, fundamental gates.
In this study, we use CNOT gates as the fundamental two-input gates, in accordance with previous studies.
It is known that the set of any single-input gates and CNOT gates is universal, meaning that any quantum gate can be formed using these gates~\cite{QCQI-book}.

Naive decomposition methods have been proposed that require $O(4^N)$ CNOT gates to decompose an arbitrary $N$-input gate~\cite{vartiainen2004efficient, shende2006synthesis}.
However, these methods are only effective for small gates due to the exponential increase in the number of CNOT gates required.
For $N=2$, a decomposition method has been proposed that requires at most 3 CNOT gates and 8 single-input gates~\cite{vidal2004universal}.

For practical use, pattern-matching methods have been proposed that reduce the required number of gates by replacing partial circuits with smaller, logically equivalent circuits~\cite{biswal2018template,dueck2018optimization,maslov2008quantum,saeedi2011synthesis}.
It is also effective to realize specific gates with fewer CNOT gates.
For example, multi-controlled Toffoli gates can be realized with $O(N^2)$ CNOT gates~\cite{saeedi2013linear} or $O(N)$ CNOT gates when a sufficient number of ancillary qubits are available~\cite{maslov2003improved}.

\subsubsection{Qubit Routing}
Qubit routing is the process of determining the mapping from logical qubits in the circuit to the physical qubits on the device at the time each quantum gate is executed.
The ideal cost function for qubit routing is the error rate in the computation result, which is difficult to estimate accurately. 

In practice, the number of noisy gates is often used as a cost function instead of the error rate.
One common approach to qubit routing is to minimize the number of SWAP gates, which is known as the Nearest Neighbor Cost~\cite{wille2016look, zulehner2018efficient, li2019tackling}.
Using a large number of SWAP gates can increase the error rate, as each SWAP gate requires three physical CNOT gates to swap the states of physically adjacent qubits.
Another approach is to minimize the total number of physical CNOT gates, including not only the cost of SWAP gates but also the cost of remote CNOT gates.

\subsubsection{Circuit Synthesis}
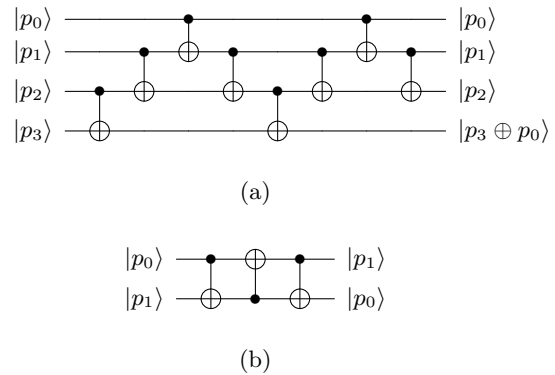
\begin{figure}[t]
    \begin{subfigure}[]{\linewidth}
        \centering
            \[ \Qcircuit @C=1em @R=0.8em {
            \lstick{\ket{p_0}} & \qw & \qw & \ctrl{1} & \qw & \qw & \qw & \ctrl{1} & \qw & \rstick{\ket{p_0}} \qw \\
            \lstick{\ket{p_1}} & \qw & \ctrl{1} & \targ & \ctrl{1} & \qw & \ctrl{1} & \targ & \ctrl{1} & \rstick{\ket{p_1}} \qw \\
            \lstick{\ket{p_2}} & \ctrl{1} & \targ & \qw & \targ & \ctrl{1} & \targ & \qw & \targ & \rstick{\ket{p_2}} \qw \\
            \lstick{\ket{p_3}} & \targ & \qw & \qw & \qw & \targ & \qw & \qw & \qw & \rstick{\ket{p_3 \oplus p_0}} \qw
            } \]
        \subcaption{}
        \label{fig:remote_CNOT_gate}
    \end{subfigure}
    \begin{subfigure}[]{\linewidth}
\centering
            \[ \Qcircuit @C=1em @R=0.8em {
            \lstick{\ket{p_0}} & \ctrl{1} & \targ & \ctrl{1} & \rstick{\ket{p_1}} \qw \\
            \lstick{\ket{p_1}} & \targ & \ctrl{-1} & \targ & \rstick{\ket{p_0}} \qw
            } \]
\subcaption{}
        \label{fig:SWAP_gate}
    \end{subfigure}
\caption{
    Decomposition of a remote CNOT gate and a SWAP gate.
    (a) Remote CNOT gate from $p_0$ to $p_3$,
    requiring 8 physical CNOT gates to implement since there are two physical qubits between them ($p_1$ and $p_2$).
    (b) SWAP gate between $p_0$ and $p_1$, requiring 3 physical CNOT gates to implement.
}
\end{figure}

In circuit synthesis, physical CNOT gates are used to build logical CNOT gates and move logical qubits.
The cost of building a logical CNOT gate increases with the distance between the control and target qubits.
In particular, the cost is only one when these qubits are adjacent, whereas the cost equals four times the number of physical qubits between the control and target qubits when they are further apart since a remote CNOT gate must be built between them~\cite{de2020quantum}.
The cost of moving logical qubits increases with the distance traveled by the logical qubits. The associated movement can be achieved using SWAP gates, which require three physical CNOT gates to swap the states of physically adjacent qubits.
Figures~\ref{fig:remote_CNOT_gate} and~\ref{fig:SWAP_gate} illustrate the decomposition of a remote CNOT gate and a SWAP gate, respectively.

\subsubsection{Example}
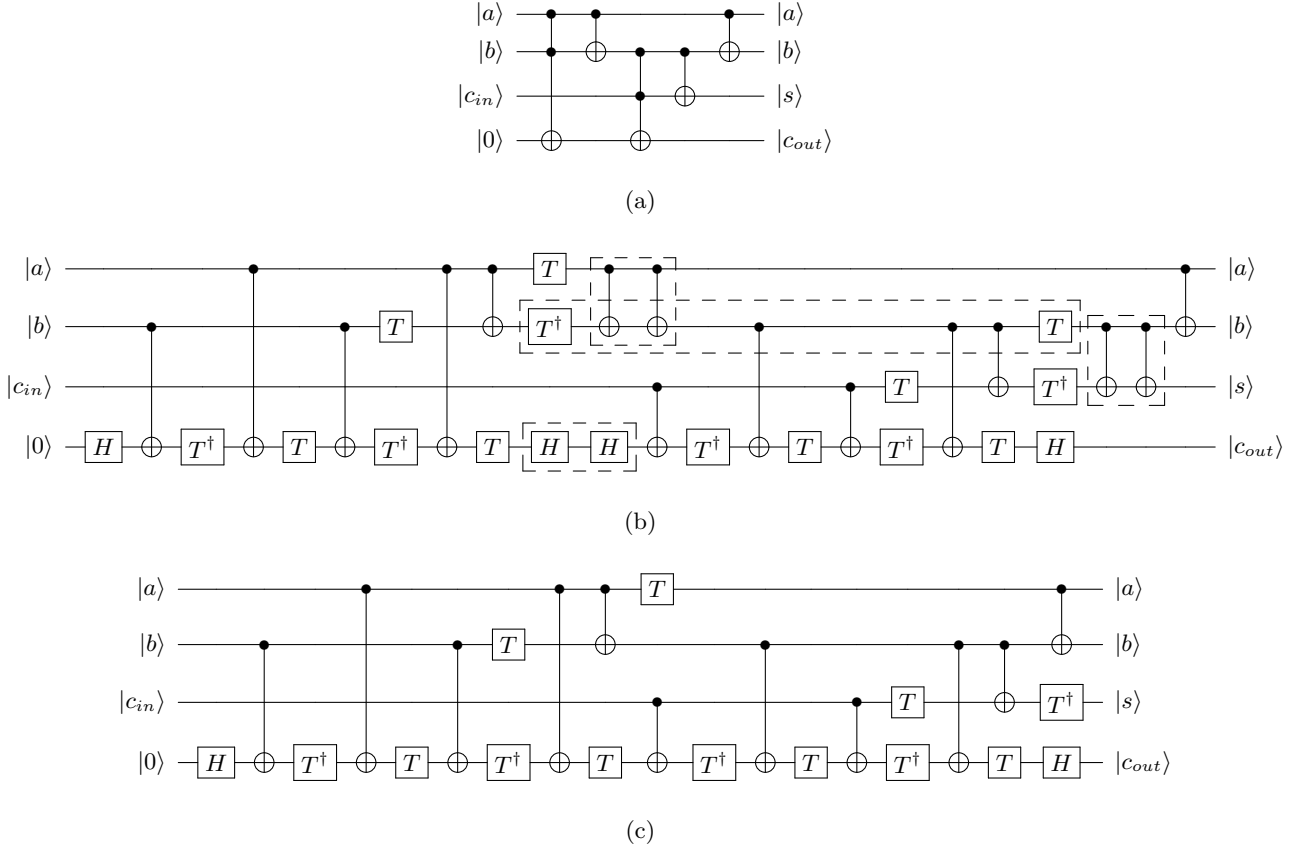
\begin{figure*}[htbp]
\centering
    \begin{subfigure}[]{\linewidth}
        \centering
        \[ \Qcircuit @C=1em @R=1em {
        \lstick{\ket{a}} & \ctrl{3} & \ctrl{1} & \qw & \qw & \ctrl{1} & \rstick{\ket{a}} \qw \\
        \lstick{\ket{b}} & \ctrl{0} & \targ & \ctrl{2} & \ctrl{1} & \targ & \rstick{\ket{b}} \qw \\
        \lstick{\ket{c_{in}}} & \qw & \qw & \ctrl{0} & \targ & \qw & \rstick{\ket{s}} \qw \\
        \lstick{\ket{0}} & \targ & \qw & \targ & \qw & \qw & \rstick{\ket{c_{out}}} \qw
        } \]
\subcaption{}
        \label{fig:example_A}
    \end{subfigure}
    
    \begin{subfigure}[]{\linewidth}
    \centering
        \[ \Qcircuit @C=0.8em @R=1em {
        \lstick{\ket{a}} & \qw & \qw & \qw & \ctrl{3} & \qw & \qw & \qw & \ctrl{3} & \ctrl{1} & \gate{T} & \ctrl{1} & \ctrl{1} & \qw & \qw & \qw & \qw & \qw & \qw & \qw & \qw & \qw & \qw & \ctrl{1} & \rstick{\ket{a}} \qw \\
        \lstick{\ket{b}} & \qw & \ctrl{2} & \qw & \qw & \qw & \ctrl{2} & \gate{T} & \qw & \targ & \gate{T^\dagger} & \targ & \targ & \qw & \ctrl{2} & \qw & \qw & \qw & \ctrl{2} & \ctrl{1} & \gate{T} & \ctrl{1} & \ctrl{1} & \targ & \rstick{\ket{b}} \qw \\
        \lstick{\ket{c_{in}}} & \qw & \qw & \qw & \qw & \qw & \qw & \qw & \qw & \qw & \qw & \qw & \ctrl{1} & \qw & \qw & \qw & \ctrl{1} & \gate{T} & \qw & \targ & \gate{T^\dagger} & \targ & \targ & \qw & \rstick{\ket{s}} \qw \\
        \lstick{\ket{0}} & \gate{H} & \targ & \gate{T^\dagger} & \targ & \gate{T} & \targ & \gate{T^\dagger} & \targ & \gate{T} & \gate{H} & \gate{H} & \targ & \gate{T^\dagger} & \targ & \gate{T} & \targ & \gate{T^\dagger} & \targ & \gate{T} & \gate{H} & \qw & \qw & \qw & \rstick{\ket{c_{out}}} \qw
        \gategroup{1}{12}{2}{13}{.7em}{--}
        \gategroup{2}{11}{2}{21}{.7em}{--}
        \gategroup{4}{11}{4}{12}{.7em}{--}
        \gategroup{2}{22}{3}{23}{.7em}{--}
        }\]
\subcaption{}
        \label{fig:example_C}
    \end{subfigure}
    
    \begin{subfigure}[]{\linewidth}
    \centering
        \[ \Qcircuit @C=0.8em @R=1em {
        \lstick{\ket{a}} & \qw & \qw & \qw & \ctrl{3} & \qw & \qw & \qw & \ctrl{3} & \ctrl{1} & \gate{T} & \qw & \qw & \qw & \qw & \qw & \qw & \qw & \ctrl{1} & \rstick{\ket{a}} \qw \\
        \lstick{\ket{b}} & \qw & \ctrl{2} & \qw & \qw & \qw & \ctrl{2} & \gate{T} & \qw & \targ & \qw & \qw & \ctrl{2} & \qw & \qw & \qw & \ctrl{2} & \ctrl{1} & \targ & \rstick{\ket{b}} \qw \\
        \lstick{\ket{c_{in}}} & \qw & \qw & \qw & \qw & \qw & \qw & \qw & \qw & \qw & \ctrl{1} & \qw & \qw & \qw & \ctrl{1} & \gate{T} & \qw & \targ & \gate{T^\dagger} & \rstick{\ket{s}} \qw \\
        \lstick{\ket{0}} & \gate{H} & \targ & \gate{T^\dagger} & \targ & \gate{T} & \targ & \gate{T^\dagger} & \targ & \gate{T} & \targ & \gate{T^\dagger} & \targ & \gate{T} & \targ & \gate{T^\dagger} & \targ & \gate{T} & \gate{H} & \rstick{\ket{c_{out}}} \qw
        } \]
\subcaption{}
        \label{fig:example_D}
    \end{subfigure}
    
    \caption{Example of quantum circuit decomposition and simplification. (a) Original circuit: A 4-qubit full-adder. (b) Decomposed circuit: Toffoli gates in the original circuit are broken down into smaller gates. (c) Simplified circuit: Pairs of canceling gates in the decomposed circuit have been removed. }
    \label{fig:example_decomposition}
\end{figure*}

We demonstrate an example of quantum circuit compilation using a 4-qubit full-adder circuit (Fig.~\ref{fig:example_A}).
First, the original gates in the circuit are decomposed into smaller gates using gate decomposition (Fig.~\ref{fig:example_C}).
This involves breaking down the two Toffoli gates into single-input gates and CNOT gates.
Next, the circuit is simplified by canceling pairs of gates (Fig.~\ref{fig:example_D}).

\begin{figure*}[htbp]
\centering
    \begin{subfigure}[]{\linewidth}
        \centering
        \includegraphics[width=0.166\linewidth]{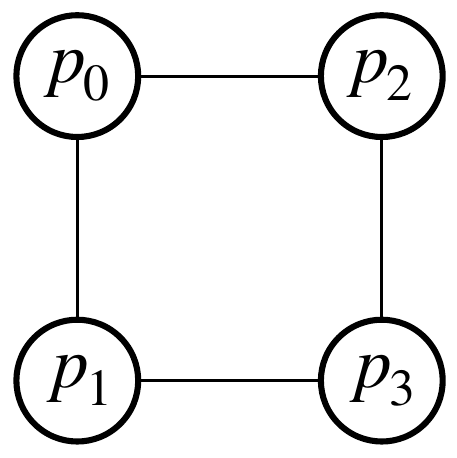}
\subcaption{}
        \label{fig:example_B}
    \end{subfigure}
    
    \begin{subfigure}[]{\linewidth}
    \centering
        \[ \Qcircuit @C=0.8em @R=1em {
        \lstick{\ket{p_0=c_{in}}} & \qw & \qw & \qw & \qw & \qw & \qw & \qw & \qw & \qswap \qwx[1] & \ctrl{2} & \gate{T} & \qw & \qw & \qw & \qw & \qw & \qw & \qw & \ctrl{2} & \qw & \rstick{\ket{p_0=a}} \qw \\
        \lstick{\ket{p_1=a}} & \qw & \qw & \qw & \ctrl{2} & \qw & \qw & \qw & \ctrl{2} & \qswap & \qw & \ctrl{2} & \qw & \qw & \qw & \ctrl{2} & \gate{T} & \qw & \targ & \qw & \gate{T^\dagger} & \rstick{\ket{p_1=s}} \qw \\
        \lstick{\ket{p_2=b}} & \qw & \ctrl{1} & \qw & \qw & \qw & \ctrl{1} & \gate{T} & \qw & \qw & \targ & \qw & \qw & \ctrl{1} & \qw & \qw & \qw & \ctrl{1} & \ctrl{-1} & \targ & \qw & \rstick{\ket{p_2=b}} \qw \\
        \lstick{\ket{p_3=0}} & \gate{H} & \targ & \gate{T^\dagger} & \targ & \gate{T} & \targ & \gate{T^\dagger} & \targ & \qw & \gate{T} & \targ & \gate{T^\dagger} & \targ & \gate{T} & \targ & \gate{T^\dagger} & \targ & \qw & \gate{T} & \gate{H} & \rstick{\ket{p_3=c_{out}}} \qw
        \gategroup{2}{18}{4}{19}{.7em}{--}
        \gategroup{1}{10}{2}{10}{1.2em}{--}
        } \]
\subcaption{}
        \label{fig:example_E}
    \end{subfigure}
    
    \begin{subfigure}[]{\linewidth}
    \centering
        \[ \Qcircuit @C=0.8em @R=1em {
        \lstick{\ket{p_0=c_{in}}} & \qw & \qw & \qw & \qw & \qw & \qw & \qw & \qw & \ctrl{1} & \targ & \ctrl{1} & \ctrl{2} & \gate{T} & \qw & \qw & \qw & \qw & \qw & \qw & \qw & \qw & \ctrl{2} & \qw & \rstick{\ket{p_0=a}} \qw \\
        \lstick{\ket{p_1=a}} & \qw & \qw & \qw & \ctrl{2} & \qw & \qw & \qw & \ctrl{2} & \targ & \ctrl{-1} & \targ & \qw & \ctrl{2} & \qw & \qw & \qw & \ctrl{2} & \gate{T} & \targ & \qw & \targ & \qw & \gate{T^\dagger} & \rstick{\ket{p_1=s}} \qw \\
        \lstick{\ket{p_2=b}} & \qw & \ctrl{1} & \qw & \qw & \qw & \ctrl{1} & \gate{T} & \qw & \qw & \qw & \qw & \targ & \qw & \qw & \ctrl{1} & \qw & \qw & \qw & \qw & \ctrl{1} & \qw & \targ & \qw & \rstick{\ket{p_2=b}} \qw \\
        \lstick{\ket{p_3=0}} & \gate{H} & \targ & \gate{T^\dagger} & \targ & \gate{T} & \targ & \gate{T^\dagger} & \targ & \qw & \qw & \qw & \gate{T} & \targ & \gate{T^\dagger} & \targ & \gate{T} & \targ & \gate{T^\dagger} & \ctrl{-2} & \targ & \ctrl{-2} & \gate{T} & \gate{H} & \rstick{\ket{p_3=c_{out}}} \qw
        \gategroup{2}{20}{4}{22}{.7em}{--}
        \gategroup{1}{10}{2}{12}{.7em}{--}
        } \]
\subcaption{}
        \label{fig:example_F}
    \end{subfigure}

\caption{Example of qubit routing and circuit synthesis. (a) Coupling graph of the physical device: Physical CNOT gates can only be applied to pairs of connected qubits. (b) Simplified circuit after placing logical qubits. (c) Synthesized circuit: All quantum gates in the simplified circuit are implemented using only physical gates.}
    \label{fig:example_compilation}
\end{figure*}

The physical qubits in this example are arranged as a 4-qubit ring (Fig.~\ref{fig:example_B}).
In qubit routing, the logical qubits are assigned to the physical qubits so that the gates can be executed efficiently (Fig.~\ref{fig:example_E}).
This circuit contains gates that require additional steps to be executed, such as a SWAP gate between $p_0$ and $p_1$, and a remote CNOT gate between $p_1$ and $p_2$.
Finally, in circuit synthesis, all gates in the circuit are implemented using only physical gates (Fig.~\ref{fig:example_F}).
The final circuit requires a total of 15 physical CNOT gates to execute.

 \section{QUBO Formulation}
\label{section:qubo_formulation}
To solve qubit mapping, ISAAQ formulates the minimization problem of  \textit{compilation cost} (i.e., the number of physical CNOT gates) as a set of QUBO problems, and solves them using Ising machines.
In this section, we first demonstrate how to represent the routes of logical qubits using binary variables.
Then, we show how to represent the compilation cost in a quadratic form.

\subsection{Routes of Logical Qubits}
To determine the routes of logical qubits, the only requirement is to determine the mapping when each logical CNOT gate is executed since single-input gates can be executed on any physical qubit without incurring additional costs.
However, assigning a mapping to each logical CNOT gate would make the QUBO model too large, increasing search time and decreasing solution quality.
To overcome this problem, ISAAQ slices the original circuit into $M$ layers and assigns a mapping to each layer to approximately represent the routes of logical qubits.

\begin{figure}[t]
\centering
    \begin{subfigure}[]{0.51\linewidth}
        \centering
        \includegraphics[width=0.9\linewidth]{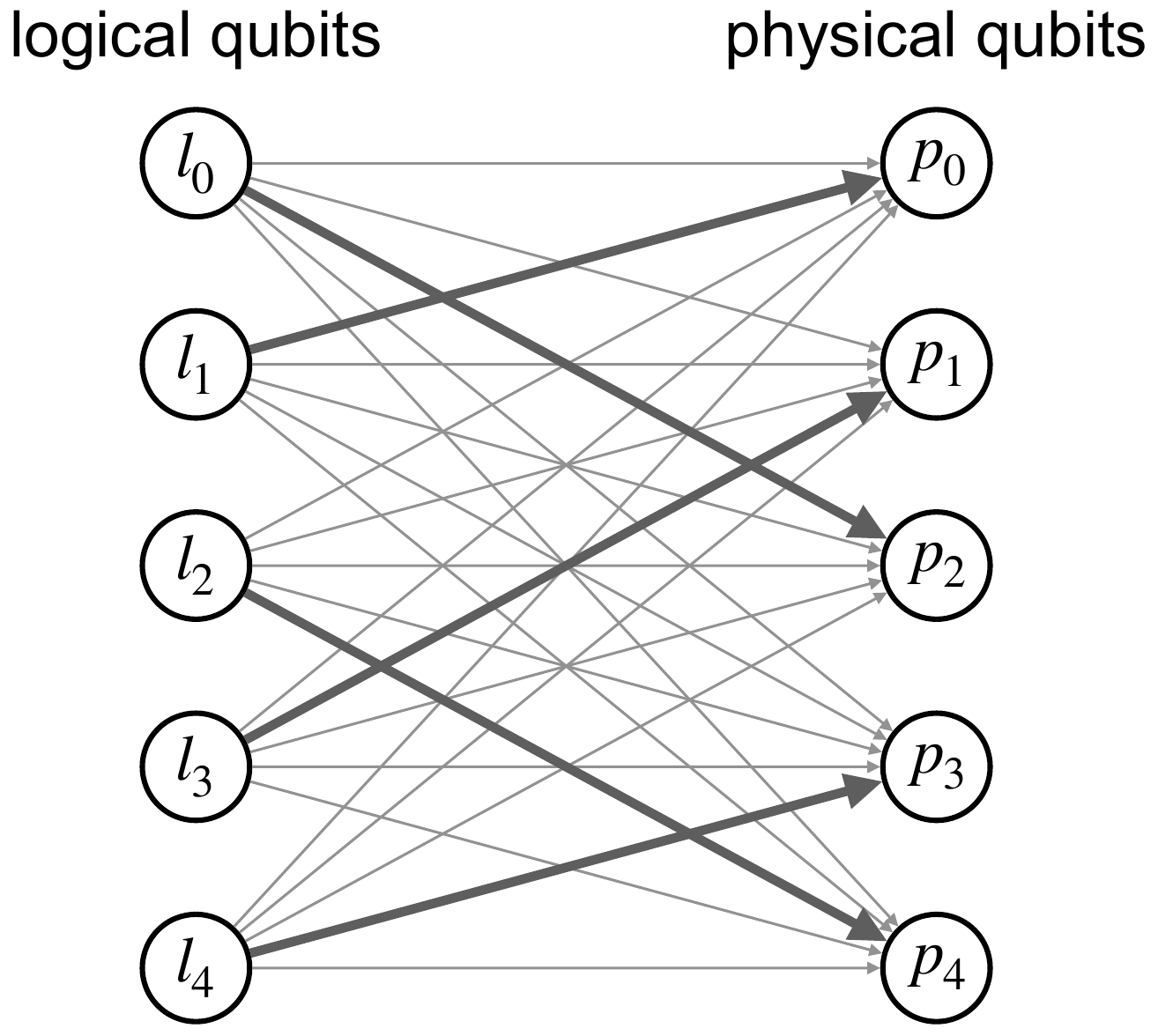}
        \subcaption{}
    \end{subfigure}
    \begin{subfigure}[]{0.47\linewidth}
        \centering
        \includegraphics[width=0.9\linewidth]{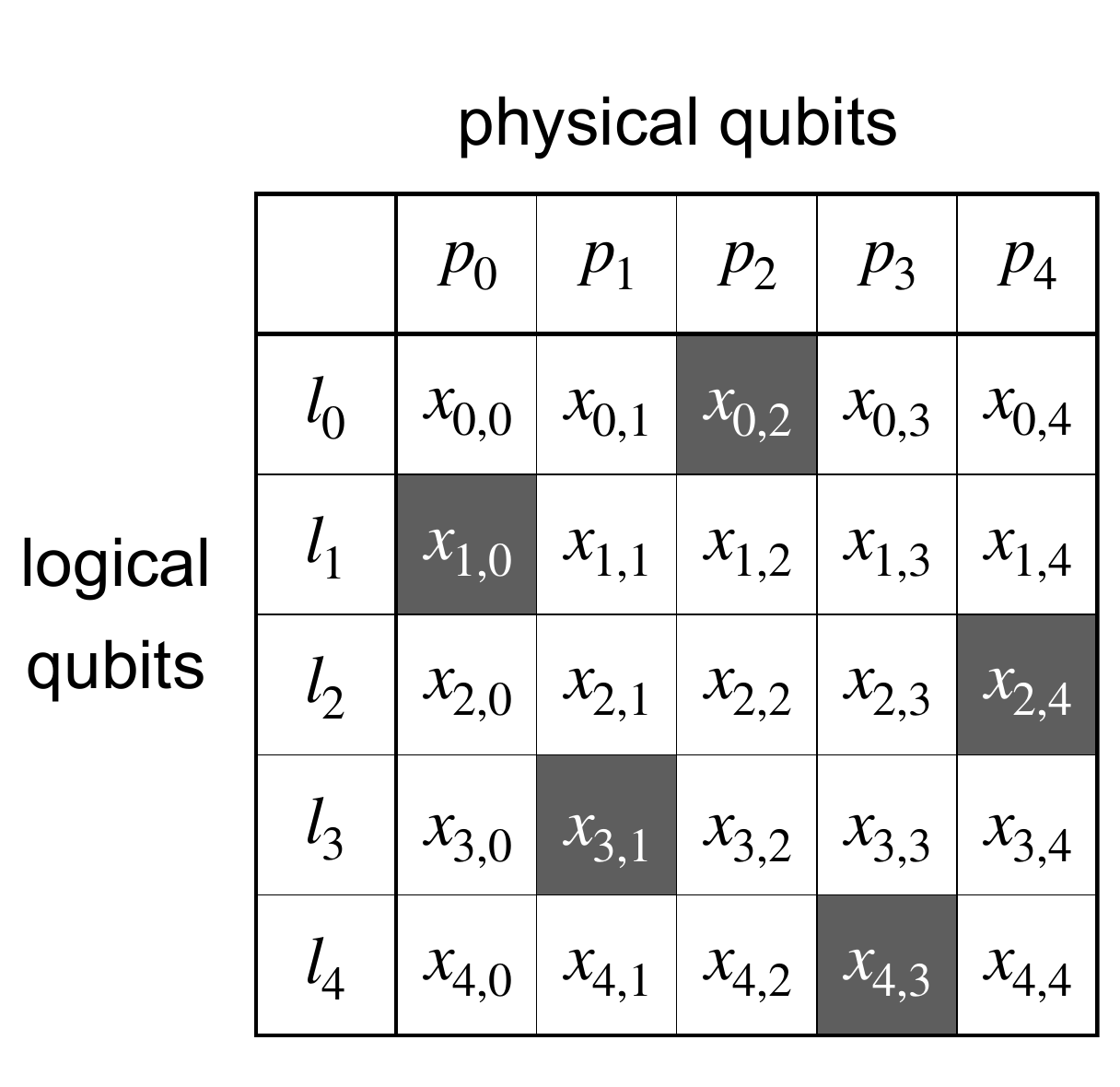}
        \subcaption{}
    \end{subfigure}
    \caption{Example of a mapping from logical qubits $l_i$ to physical qubits $p_\mu$. (a) Visual representation of the mapping as a bijective function $f^m$. (b) Corresponding state of binary variables $x^m_{i, \mu}$, where the shaded cells and the non-shaded cells represent 1 and 0, respectively.}
    \label{fig:binary_variables_mapping}
\end{figure}

The qubit routing problem can be formulated as a problem of finding a sequence consisting of $M$ mappings. The mapping of logical qubits in the $m$-th layer, denoted by $\bm{p^m}$, can be formulated as follows:
\begin{align}
\bm{p^m}
&= \begin{pmatrix}
p^m_0 & \cdots & p^m_{N-1} \\
\end{pmatrix} \nonumber \\
&\coloneqq \begin{pmatrix}
f^m(l_0) & \cdots & f^m(l_{N-1}) \\
\end{pmatrix},
\end{align}
where $f^m$ is a bijective function that connects logical and physical qubits.
In the QUBO formulation, we use binary variables to represent whether an edge is drawn or not.
Specifically, we assign a binary variable $x^m_{i,\mu}$ to each edge, defined as $x^m_{i,\mu} \coloneqq \bm{1}\left( p^m_i = p_\mu \right)$.
Figure~\ref{fig:binary_variables_mapping} shows an example of $f^m$ and $x^m_{i, \mu}$, where the logical qubits $(l_0, l_1, l_2, l_3, l_4)$ are assigned to $(p_2, p_0, p_4, p_1, p_3)$, respectively.

To represent a bijection, we use the one-hot constraint for each row and column as shown below:
\begin{align}
\sum_{i=0}^{N-1} x^m_{i,\mu} = \sum_{\mu=0}^{N-1} x^m_{i,\mu} = 1.
\end{align}
Since QUBO problems cannot handle constraints directly, the constraints are converted to penalty terms and added to the QUBO cost as follows:
\begin{align}
\mathrm{penalty} = \lambda\left(\sum_{i=0}^{N-1} x^m_{i,\mu} - 1\right)^2 + \lambda\left(\sum_{\mu=0}^{N-1} x^m_{i,\mu} - 1\right)^2,
\end{align}
where $\lambda$ is the parameter that represents the strength of a constraint.

\subsection{Compilation Cost}
We define the compilation cost as the number of physical CNOT gates.
The compilation cost consists of the \textit{building cost} and the \textit{moving cost}. These two costs represent the number of physical CNOT gates required for building the original gates and for moving logical qubits, respectively.
This approach can be formally described as follows:
\begin{align}
    \mathrm{COST}_\mathrm{cmp} = \mathrm{COST}_\mathrm{bld} + \mathrm{COST}_\mathrm{mov}.
\end{align}

The building cost is the number of physical CNOT gates required to build logical CNOT gates inside circuit layers.
For simplicity, we assume that the logical CNOT gates are built one at a time.
This assumption allows the expression of the building cost as the sum of the costs associated with each of the logical CNOT gates, formally expressed as
\begin{align}
\mathrm{COST}_\mathrm{bld}
&= \sum_{m = 0}^{M-1} \sum_{(l_c, l_t) \in G^m} c\left( f^m(l_c), f^m(l_t) \right) \nonumber \\
&= \sum_{m = 0}^{M-1} \sum_{(l_c, l_t) \in G^m} c\left( p^m_c, p^m_t \right).
\label{eq:building_cost}
\end{align}
In this equation, $G^m$ represents the set of logical CNOT gates in the $m$-th layer and $c\left( p^m_c, p^m_t \right)$ represents the cost of building a logical CNOT gate from $p^m_c$ to $p^m_t$.
The coefficients $c\left( p^m_c, p^m_t \right)$ can be determined based on the fact that a logical CNOT gate from $p_\mu$ to $p_\nu$ can be implemented using $\max\left(1, 4(d(p_\mu, p_\nu) - 1)\right)$ physical CNOT gates, as reported in~\cite{de2020quantum}.
In this expression, $d(p_\mu, p_\nu)$ represents the distance between $p_\mu$ and $p_\nu$.
By calculating $c\left( p_\mu, p_\nu \right)$ for all pairs of physical qubits in advance, we can use the following relation
\begin{align}
c\left( p^m_c, p^m_t \right)
&= \sum_{\mu = 0}^{N-1} \sum_{\nu = 0}^{N-1} c(p_\mu, p_\nu) x^m_{c,\mu} x^m_{t,\nu}
\end{align}
to formulate the building cost as
\begin{align}
\mathrm{COST}_\mathrm{bld} = \sum_{m = 0}^{M-1} \sum_{(l_c, l_t) \in G^m} \sum_{\mu = 0}^{N-1} \sum_{\nu = 0}^{N-1} c(p_\mu, p_\nu) x^m_{c,\mu} x^m_{t,\nu}.
\end{align}

\begin{figure}[t]
\centering
    \begin{subfigure}[]{0.49\linewidth}
        \centering
        \includegraphics[width=0.9\linewidth]{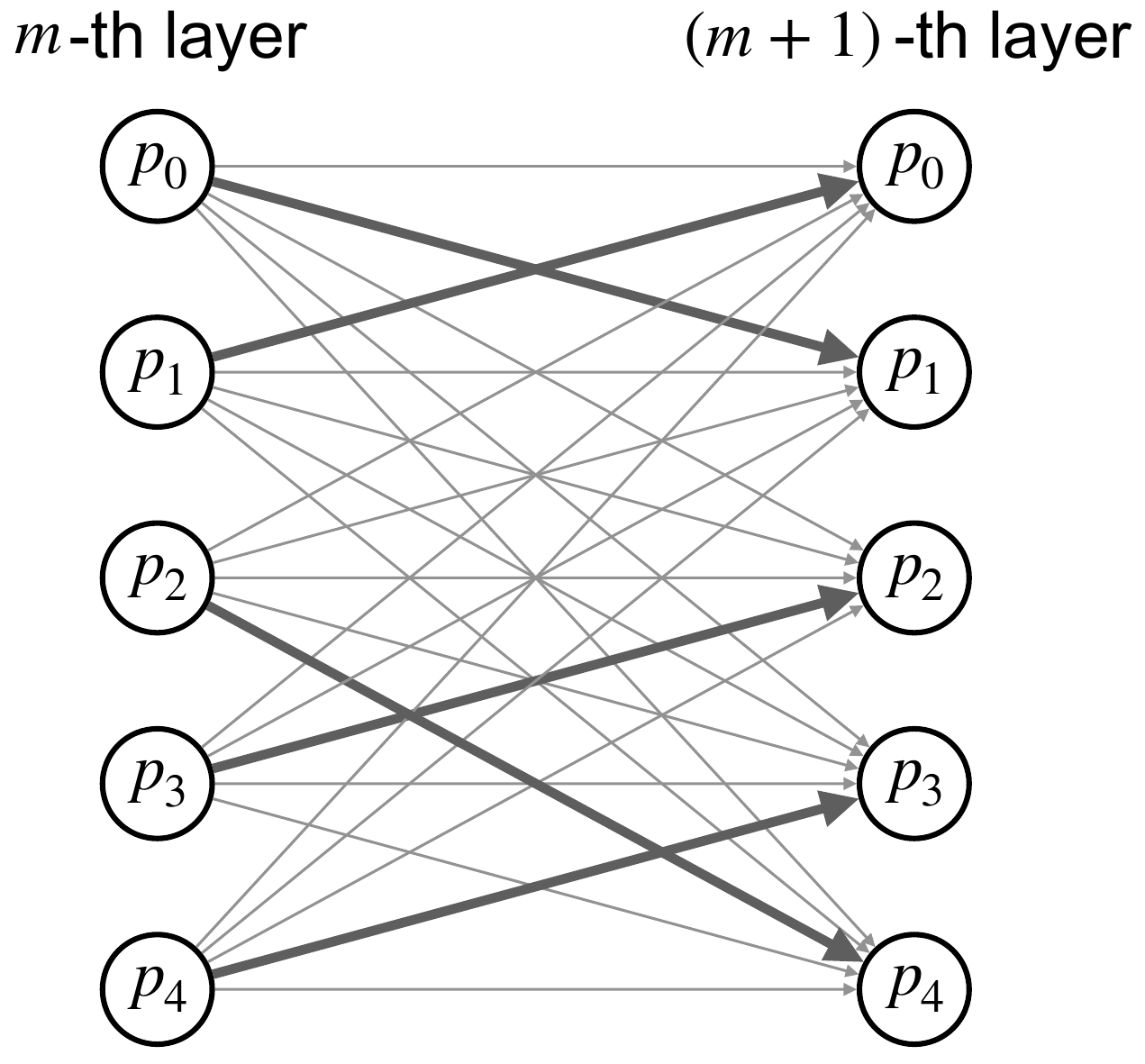}
        \subcaption{}
    \end{subfigure}
    \begin{subfigure}[]{0.49\linewidth}
        \centering
        \includegraphics[width=0.9\linewidth]{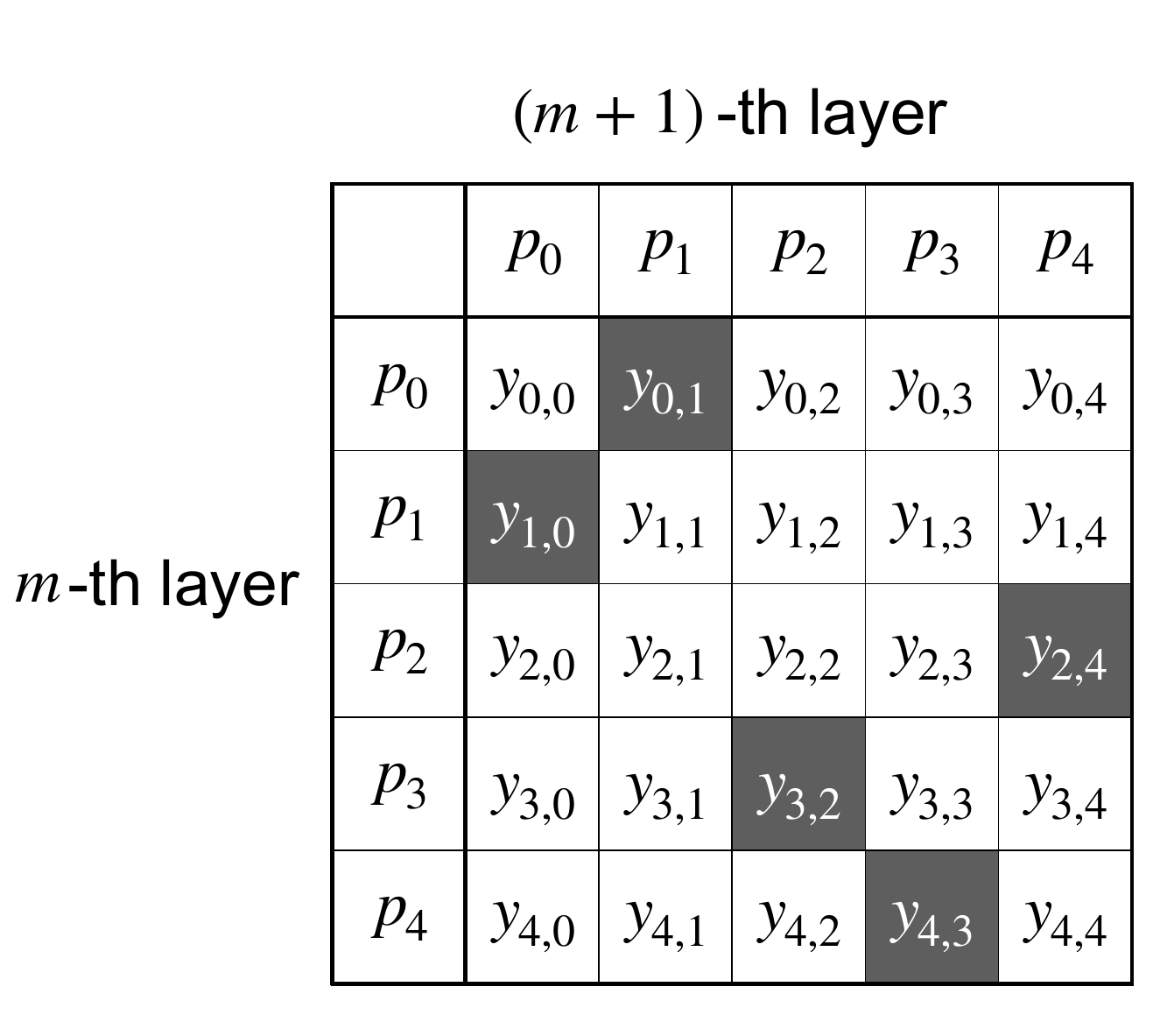}
        \subcaption{}
    \end{subfigure}
    \caption{Example of a rearrangement of logical qubits between the $m$-th layer and $(m+1)$-th layer.
    (a) Visual representation of the rearrangement $\bm{\pi^m}$. (b) Corresponding state of binary variables $y^m_{i, \mu}$, where the shaded cells and the non-shaded cells represent 1 and 0, respectively.}
    \label{fig:binary_variables_permutation}
\end{figure}

The moving cost is the number of physical CNOT gates required to rearrange the placement of logical qubits between circuit layers.
These rearrangements can be achieved using SWAP gates, which require three physical CNOT gates each.
Therefore, the moving cost can be expressed as
\begin{align}
\mathrm{COST}_\mathrm{mov} = 3\sum_{m=0}^{M-2} \mathrm{N_s}(\bm{\pi^m}),
\end{align}
where $\bm{\pi^m} = \bm{p^{m+1}}(\bm{p^m})^{-1}$ represents the rearrangement of logical qubits from the placement in $\bm{p^m}$ to the placement in $\bm{p^{m+1}}$, and $\mathrm{N_s}(\bm{\pi^m})$ represents the number of SWAP gates required to perform this rearrangement.
We define a binary variable $y^m_{\mu, \nu} \coloneqq \sum_{i=0}^{N-1} x^m_{i, \mu} x^{m+1}_{i, \nu}$ for each pair of physical qubits to represent whether a logical qubit moves from $p_\mu$ to $p_\nu$ or not.
From the definition of $y^m_{\mu, \nu}$, it can be seen that $y^m_{\mu, \nu} = \bm{1}\left( \pi^m_\mu = \nu \right)$ holds.
Figure \ref{fig:binary_variables_permutation} displays an example of $\bm{\pi^m}$ and $y^m_{i, \mu}$, where the physical qubits $(p_0, p_1, p_2, p_3, p_4)$ are rearranged to $(p_1, p_0, p_4, p_2, p_3)$.

Finding the minimum number of SWAP gates required for a given rearrangement is known as a token swapping problem, which has been proven to be NP-Hard~\cite{aichholzer2021hardness}.
This means that $\mathrm{N_s}(\bm{\pi^m})$ cannot be converted directly to a quadratic form.
To overcome this issue, we use an approximation that expresses the number of required SWAP gates in a quadratic form, as follows:
\begin{align}
\mathrm{N_s}(\bm{\pi^m})
&\approx \sum_{\mu=0}^{N-1} a_{\mu, \pi^m_\mu} \nonumber \\
&= \sum_{\mu=0}^{N-1} \sum_{\nu=0}^{N-1} a_{\mu, \nu} y^m_{\mu, \nu} \nonumber \\
&= \sum_{\mu=0}^{N-1} \sum_{\nu=0}^{N-1} a_{\mu, \nu} \sum_{i=0}^{N-1} x^m_{i, \mu} x^{m+1}_{i, \nu}.
\label{eq:SWAP_cost_approx}
\end{align}
To minimize the squared error of this approximation, we optimize the coefficients $a_{\mu, \nu}$ based on the distribution of $\bm{\pi^m}$.
We estimate this distribution using previous compilation results; consequently, running ISAAQ more frequently will improve the accuracy of the approximation.
In the case where there is no compilation result, we assume that the distribution of $\bm{\pi^m}$ follows the uniform distribution of permutations.
Further information on finding appropriate coefficients $a_{\mu, \nu}$ can be found in appendix \ref{section:determining_qubo_coefficients}.

Using these equations, the compilation cost can be formulated as follows:
\begin{align}
    \mathrm{COST}_\mathrm{cmp}
    &= \mathrm{COST}_\mathrm{bld} + \mathrm{COST}_\mathrm{mov} \nonumber \\
    &\approx \sum_{m = 0}^{M-1} \sum_{(l_c, l_t) \in G^m} \sum_{\mu = 0}^{N-1} \sum_{\nu = 0}^{N-1} c(p_\mu, p_\nu) x^m_{c,\mu} x^m_{t,\nu} \nonumber \\
    &+ 3\sum_{m=0}^{M-2} \sum_{\mu=0}^{N-1} \sum_{\nu=0}^{N-1} a_{\mu, \nu} \sum_{i=0}^{N-1} x^m_{i, \mu} x^{m+1}_{i, \nu}.
\end{align}

 \section{ISAAQ}
\label{section:isaaq}

\subsection{System Overview}
ISAAQ involves three steps for quantum circuit compilation: circuit partitioning, qubit routing, and circuit synthesis.
In the circuit partitioning step, ISAAQ divides the original circuit into \textit{circuit layers} based on the number of logical CNOT gates.
Circuit layers are then combined into several \textit{circuit chunks} based on the capabilities of Ising machines.
In the qubit routing step, ISAAQ creates a QUBO problem for each chunk to represent the compilation cost and performs qubit routing by solving these QUBO problems using Ising machines.
This step can be parallelized when multiple Ising machines are available.
Finally, in the circuit synthesis step, ISAAQ builds the original gates based on the placement of logical qubits and rearranges these qubits using SWAP gates between circuit layers.
After that, ISAAQ updates the coefficients of its QUBO model using the rearrangements obtained from the compilation result.

\subsection{Circuit Chunk Scheduling}
\label{section:circuit_chunk_scheduling}
During qubit routing, ISAAQ solves multiple QUBO problems in parallel by creating a schedule for determining the order in which to solve them.
However, there is a trade-off between minimizing the compilation cost and minimizing the compilation time.

To reduce the compilation cost, it is necessary to include the moving cost in the QUBO problems.
This facilitates the avoidance of quasi-random rearrangements that would require a large number of physical CNOT gates.
However, since each QUBO problem can only optimize the placement of logical qubits within a single chunk, the moving cost between chunks cannot be included in the QUBO problems until one of the problems has been solved.
Therefore, it is not advisable to solve the QUBO problems for neighboring chunks simultaneously.
Conversely, minimizing the compilation time can be achieved by using multiple Ising machines in parallel to solve the QUBO problems.
This strategy significantly speeds up the process of solving the QUBO problems.
Therefore, it is important to keep many Ising machines running by reducing the number of QUBO problems that are dependent on the results of others.

There are two simple scheduling strategies that can be used to solve multiple QUBO problems with multiple Ising machines: the \textit{independent strategy} and the \textit{sequential strategy}.
In this section, the number of QUBO problems and the number of Ising machines are represented as $Q$ and $I$, respectively.

\begin{figure}[t]
    \begin{subfigure}[]{\linewidth}
        \centering
        \includegraphics[width=\linewidth]{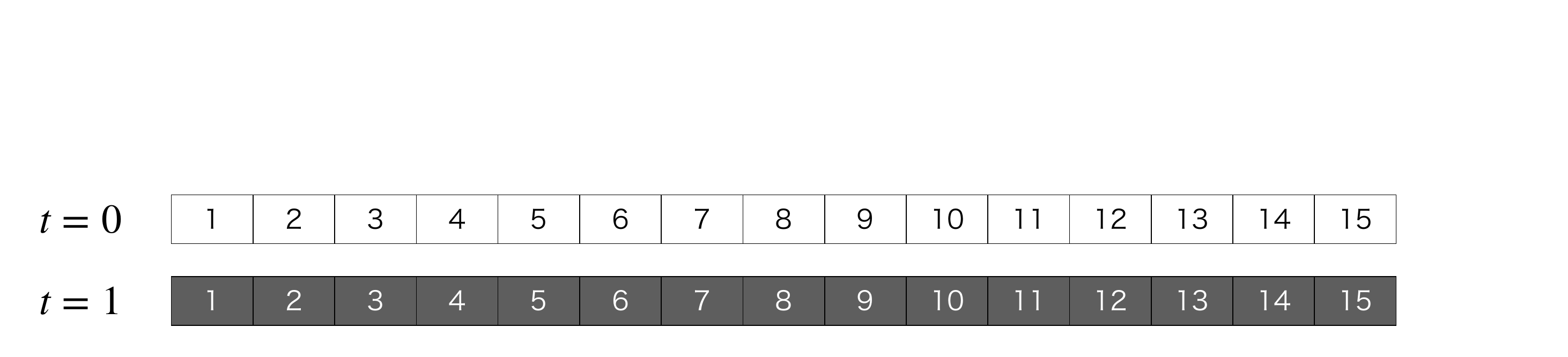}
        \subcaption{}
        \label{fig:chunks_a}
    \end{subfigure}
    \begin{subfigure}[]{\linewidth}
        \centering
        \includegraphics[width=\linewidth]{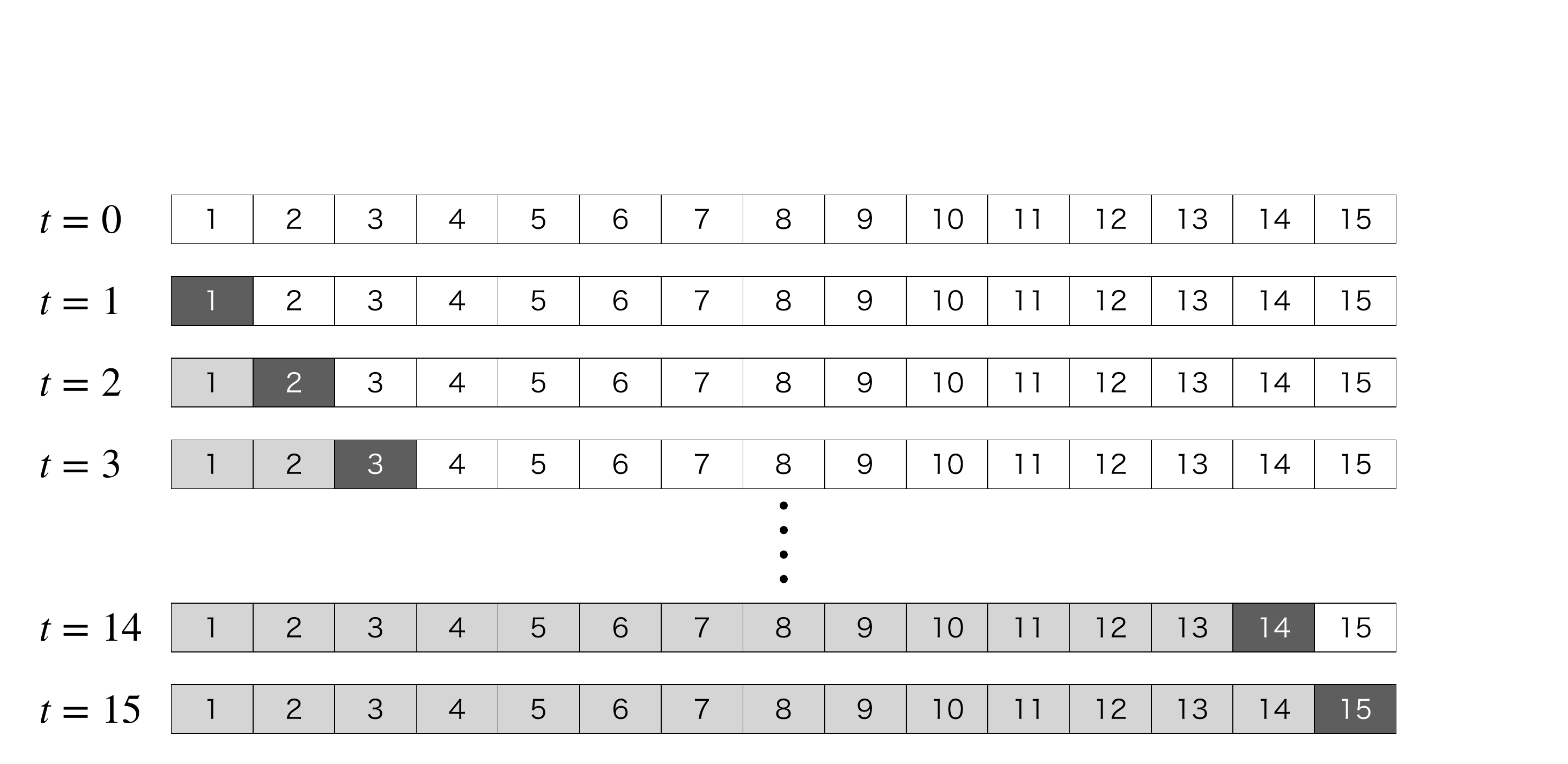}
        \subcaption{}
        \label{fig:chunks_b}
    \end{subfigure}
    \begin{subfigure}[]{\linewidth}
        \centering
        \includegraphics[width=\linewidth]{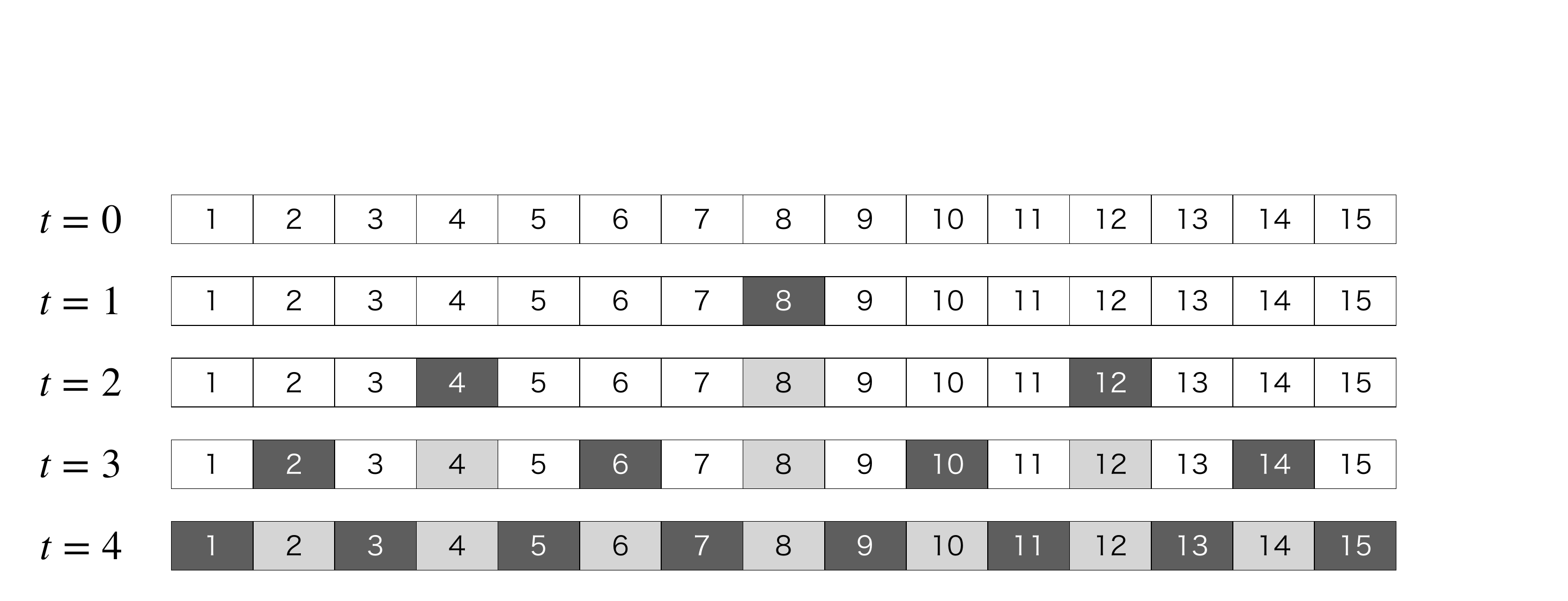}
        \subcaption{}
        \label{fig:chunks_c}
    \end{subfigure}
    \caption{
Comparison of scheduling strategies for solving 15 QUBO problems.
    The white, inverted, and shaded cells represent the QUBO problems remaining to be solved, in progress, and already solved, respectively.
    (a) Independent strategy. (b) Sequential Strategy. (c) Binary strategy.
    }
\end{figure}

The independent strategy, illustrated in Fig.~\ref{fig:chunks_a}, allows all QUBO problems to be solved in parallel, making the most efficient use of Ising machines.
This strategy takes only $\ceil{Q/I}$ time steps to solve all QUBO problems using this strategy.
However, it does not consider the moving costs between circuit chunks.
The sequential strategy, illustrated in Fig.~\ref{fig:chunks_b}, involves solving the QUBO problems from the left-most to the right-most one.
This strategy allows the moving costs between circuit chunks to be included in the QUBO problems.
However, it takes $Q$ time steps to solve all problems regardless of the number of Ising machines available.
Overall, both the independent and sequential strategies have their respective advantages and disadvantages.
The independent strategy is quicker but does not consider the moving costs, whereas the sequential strategy does consider the moving costs but is slower.

In this study, we propose a new strategy named the \textit{binary strategy}, which combines the acceleration of the process with the minimization of the moving costs.
As shown in Fig.~\ref{fig:chunks_c}, the QUBO problems are placed on a binary tree and solved from the root to the leaves.
Each QUBO problem includes the moving cost between the nearest processed chunks.
Although the moving costs may take the form of virtual terms when the nearest processed chunk is far away, they have the effect of minimizing the total moving cost.
To reflect the importance of these virtual cost terms, they are weighted by a factor of $1/d$, where $d$ is the distance between the nearest chunks.

Using the binary strategy, the QUBO problems placed at the same height on the binary tree can be solved in parallel because they depend only on the roots.
The total number of time steps required using this strategy can be calculated by summing the time steps for each height as follows:
\begin{align}
& \phantom{{}<{}} \sum_{i=0}^{\floor{\log_2{Q}}} \ceil{\frac{1}{I}\left(\floor{\frac{Q}{2^i}} - \floor{\frac{Q}{2^{i+1}}}\right)} \nonumber \\
& < \sum_{i=0}^{\floor{\log_2{Q}}} \left(1 + \frac{1}{I}\floor{\frac{Q}{2^i}} - \frac{1}{I}\floor{\frac{Q}{2^{i+1}}}\right) \nonumber \\
& = \frac{Q}{I} + \floor{\log_2{Q}} + 1.
\end{align}
This result indicates that the binary strategy requires at most $\floor{\log_2{Q}} + 1$ time steps more than the independent strategy while still allowing the QUBO problems to include the moving costs between circuit chunks.

\subsection{Circuit synthesis}
Based on the routing result, ISAAQ synthesizes a circuit that is logically equivalent to the original circuit, using only physical quantum gates.
This process consists of two steps: \textit{circuit layer construction} and \textit{circuit layer stitching}.
In the circuit layer construction step, ISAAQ implements the original gates within the circuit layers.
In the circuit layer stitching step, ISAAQ moves logical qubits to rearrange their placement between circuit layers.

\subsubsection{Circuit layer construction}
After determining the placement of logical qubits in each circuit layer, ISAAQ implements the original gates that act on these qubits.
Physical CNOT gates are not required to implement the original single-input gates because of the assumption that single-input gates can be executed on any physical qubit.
Conversely, the implementation of logical CNOT gates that act on non-adjacent physical qubits requires additional physical CNOT gates, as shown in Fig.~\ref{fig:remote_CNOT_gate}.

\begin{figure}[t]
\centering
    \begin{subfigure}[]{0.49\linewidth}
        \centering
        \[ \Qcircuit @C=1em @R=1em {
        \lstick{\ket{p_0}} & \ctrl{2} & \ctrl{3} & \ctrl{4} & \qw \\
        \lstick{\ket{p_1}} & \qw & \qw & \qw & \qw \\
        \lstick{\ket{p_2}} & \targ & \qw & \qw & \qw \\
        \lstick{\ket{p_3}} & \qw & \targ & \qw & \qw \\
        \lstick{\ket{p_4}} & \qw & \qw & \targ & \qw
        } \]
        \subcaption{}
        \label{fig:hub_example_a}
    \end{subfigure}
    \begin{subfigure}[]{0.49\linewidth}
        \centering
        \vspace{6pt}
        \includegraphics[width=\linewidth]{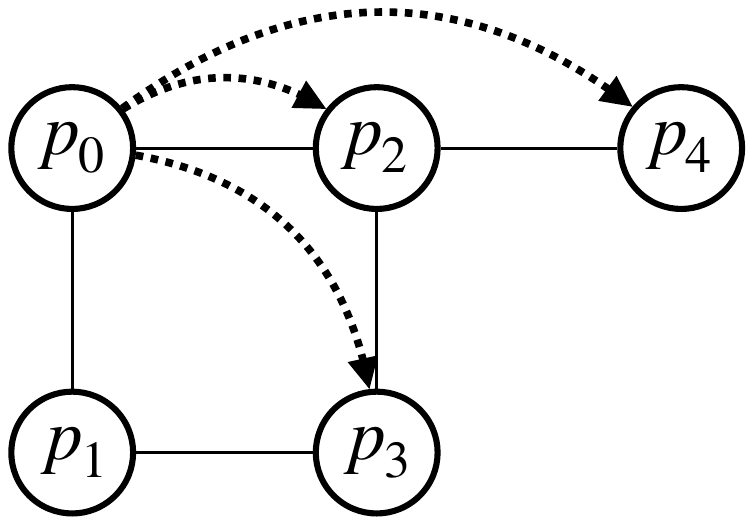}
        \subcaption{}
        \label{fig:hub_example_b}
    \end{subfigure}
    \begin{subfigure}[]{\linewidth}
        \centering
        \[ \Qcircuit @C=1em @R=1em {
        \lstick{\ket{p_0}} & \ctrl{2} & \qw & \ctrl{2} & \qw & \ctrl{2} & \qw & \ctrl{2} & \qw & \ctrl{2} & \qw \\
        \lstick{\ket{p_1}} & \qw & \qw & \qw & \qw & \qw & \qw & \qw & \qw & \qw & \qw \\
        \lstick{\ket{p_2}} & \targ & \ctrl{1} & \targ & \ctrl{1} & \targ & \ctrl{2} & \targ & \ctrl{2} & \targ & \qw \\
        \lstick{\ket{p_3}} & \qw & \targ & \qw & \targ & \qw & \qw & \qw & \qw & \qw & \qw \\
        \lstick{\ket{p_4}} & \qw & \qw & \qw & \qw & \qw & \targ & \qw & \targ & \qw & \qw
        } \]
        \subcaption{}
        \label{fig:hub_example_c}
    \end{subfigure}
    \begin{subfigure}[]{\linewidth}
        \centering
        \[ \Qcircuit @C=1em @R=1em {
        \lstick{\ket{p_0}} & \qw & \qw & \ctrl{2} & \qw & \qw & \qw \\
        \lstick{\ket{p_1}} & \qw & \qw & \qw & \qw & \qw & \qw \\
        \lstick{\ket{p_2}} & \ctrl{1} & \ctrl{2} & \targ & \ctrl{1} & \ctrl{2} & \qw \\
        \lstick{\ket{p_3}} & \targ & \qw & \qw & \targ & \qw & \qw \\
        \lstick{\ket{p_4}} & \qw & \targ & \qw & \qw & \targ & \qw
        } \]
        \subcaption{}
        \label{fig:hub_example_d}
    \end{subfigure}
    \caption{Example implementations of multiple CNOT gates. These logical CNOT gates share $p_0$ as the common control qubit. (a) Logical CNOT gates. (b) Physical device. (c) Implementation of the logical CNOT gates by the straightforward method. (d) Implementation of the logical CNOT gates by the cost-reduction method using $p_2$ as a relay qubit.}
    \label{fig:hub_example}
\end{figure}

The most straightforward way of synthesizing a physical circuit is to implement the original gates one by one, which requires the same number of physical CNOT gates as the building cost (described in Eq.~\eqref{eq:building_cost}).
Figure~\ref{fig:hub_example_c} shows an example of a construction result of the straightforward method, where the logical CNOT gates are implemented using 9 physical CNOT gates.
However, this approach is not optimal as we can reduce the number of physical CNOT gates by eliminating duplicates.

To optimize the number of physical CNOT gates used in the implementation of logical CNOT gates, ISAAQ employs \textit{relay qubits}, which act as caches by temporarily storing intermediate calculation results for efficient access. 
ISAAQ searches for sets of commutative logical CNOT gates that share a single physical qubit as the control or target qubit.
For example, the three logical CNOT gates displayed in Fig.~\ref{fig:hub_example_a} share $p_0$ as their control qubit.
Using $p_2$ as a relay qubit renders access to the common control qubit more efficient, allowing ISAAQ to implement the logical CNOT gates using only 5 physical CNOT gates as shown in Fig.~\ref{fig:hub_example_d}.

When ISAAQ identifies a set of commutative logical CNOT gates that share a common control or target qubit, ISAAQ compares the outcomes of two strategies for implementing these gates with the goal of minimizing the required number of physical CNOT gates.
The first strategy involves implementing the logical CNOT gates directly, without using any relay qubits.
In this case, the total number of physical CNOT gates required is given by 
\begin{align}
\sum_{p_i \in P} c(p_c, p_i) = \sum_{p_i \in P} \min(1, 4d(p_c, p_i) - 4),
\end{align}
where $p_c$ and $P$ represent the common qubit and the qubits that interact with $p_c$, respectively.
The second strategy involves the use of a relay qubit $p_r$ to facilitate the implementation of the logical CNOT gates.
The required number of physical CNOT gates in this case is given by
\begin{align}
    c(p_c, p_r) + \sum_{p_i \in P} \min(c(p_c, p_i), 4d(p_r, p_i) - 2)
\end{align}
when $p_r \in P$ and
\begin{align}
    4d(p_c, p_r) - 2 + \sum_{p_i \in P} \min(c(p_c, p_i), 4d(p_r, p_i) - 2)
\end{align}
when $p_r \notin P$.
The relay qubit $p_r$ is chosen to minimize the required number of physical CNOT gates.

\subsubsection{Circuit layer stitching}
\label{section:SWAP_insertion}
In the circuit layer stitching step, ISAAQ rearranges the placement of logical qubits between circuit layers.
To perform this action efficiently, ISAAQ attempts to minimize the number of SWAP gates used, each consisting of three physical CNOT gates.
This optimization problem, known as the token swapping problem, has been proven to be NP-Hard for general hardware topologies.

\begin{figure}[t]
    \centering
    \includegraphics[width=\linewidth]{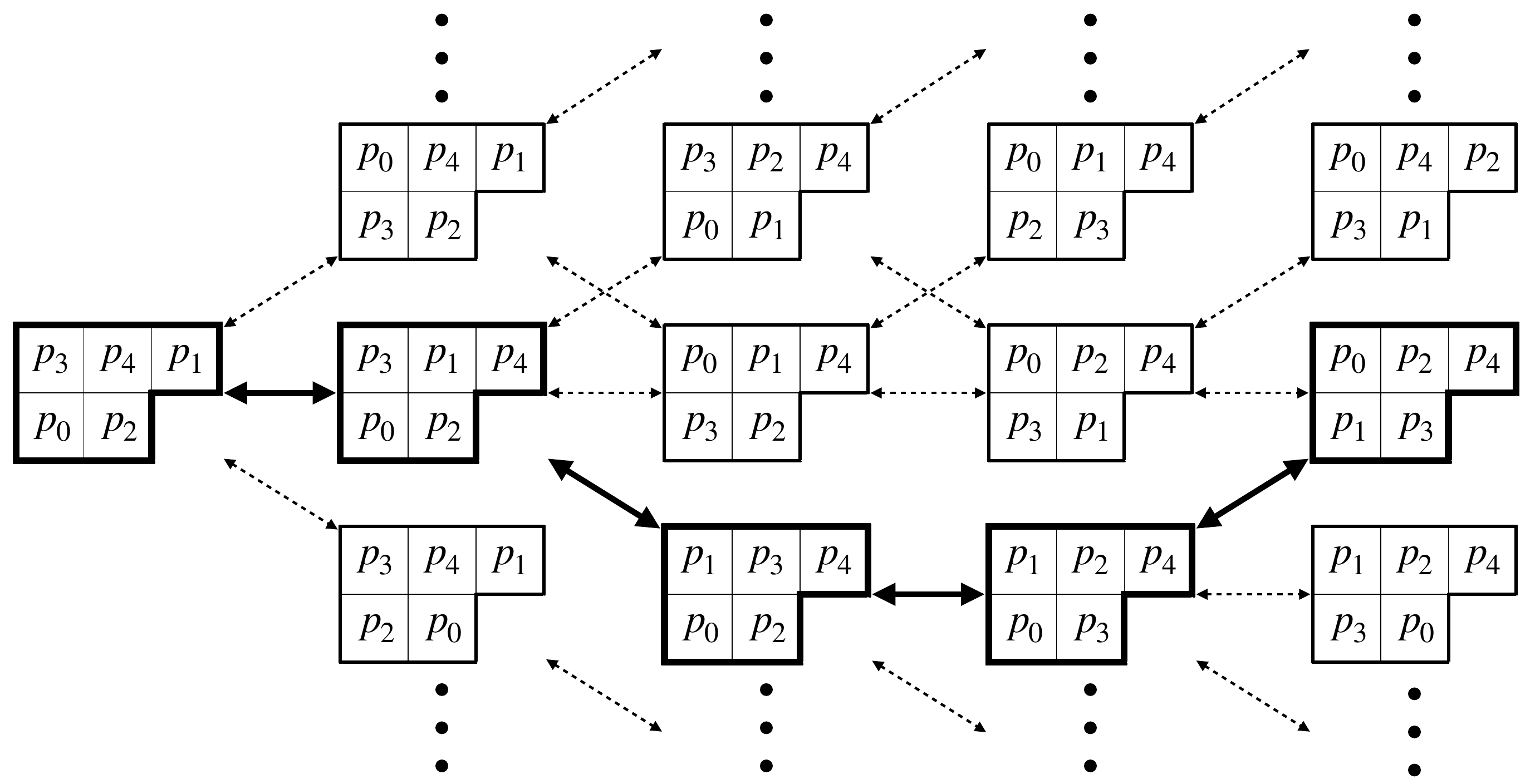}
    \caption{
    Exact solution for the token swapping problem. ISAAQ obtains the shortest sequence of SWAP gates by calculating the shortest path on the graph of permutations.
    }
    \label{fig:routing_exact}
\end{figure}

To address this challenge, we develop two solutions: an exact solution for small-scale devices and a heuristic solution for intermediate-scale devices.
The exact solution involves constructing a graph of permutations, where each node represents a placement of logical qubits and each edge represents a transition that can be achieved using a single SWAP gate.
ISAAQ uses a breadth-first search to find the shortest path on this graph, as shown in Fig.~\ref{fig:routing_exact}.
However, this solution is only effective for devices with 10 or fewer qubits, due to the exponential time required for execution.

To overcome the weakness of the exact solution, we propose the heuristic solution to efficiently identify near-optimal instructions for rearranging logical qubits.
The heuristic solution involves moving logical qubits one by one and fixing their positions once they have reached their destinations.
However, the order in which the logical qubits are moved cannot be arbitrary, because fixing a logical qubit's position may disconnect the coupling graph and prevent several logical qubits from reaching their destinations.
To determine a valid order for moving the logical qubits, ISAAQ first creates a sequence of physical qubits by iteratively choosing physical qubits from the neighbors of the previously chosen qubits and then reversing the sequence.
The reversed sequence can be used as a sequence of destinations, as it ensures that fixing the positions of the logical qubits will keep the coupling graph connected.

After establishing the order of the destinations, ISAAQ moves each logical qubit along the shortest path on the coupling graph.
In order to rearrange the logical qubits efficiently, each edge $p_\mu \rightarrow p_\nu$ is weighed by
\begin{align}
w(p_\mu \rightarrow p_\nu) &\coloneqq \Delta(p_\mu \rightarrow p_\nu; p'_s) - \Delta(p_\mu \rightarrow p_\nu; p'_\nu) + 2 \nonumber \\
\Delta(p_\mu \rightarrow p_\nu; p'_s) &\coloneqq d(p_\nu, p'_s) - d(p_\mu, p'_s),
\end{align}
where $p'_s$ and $p'_\nu$ represent the destinations of the logical qubits placed on $p_s$ and $p_\nu$, respectively.

\begin{figure}[t]
    \centering
    \includegraphics[width=\linewidth]{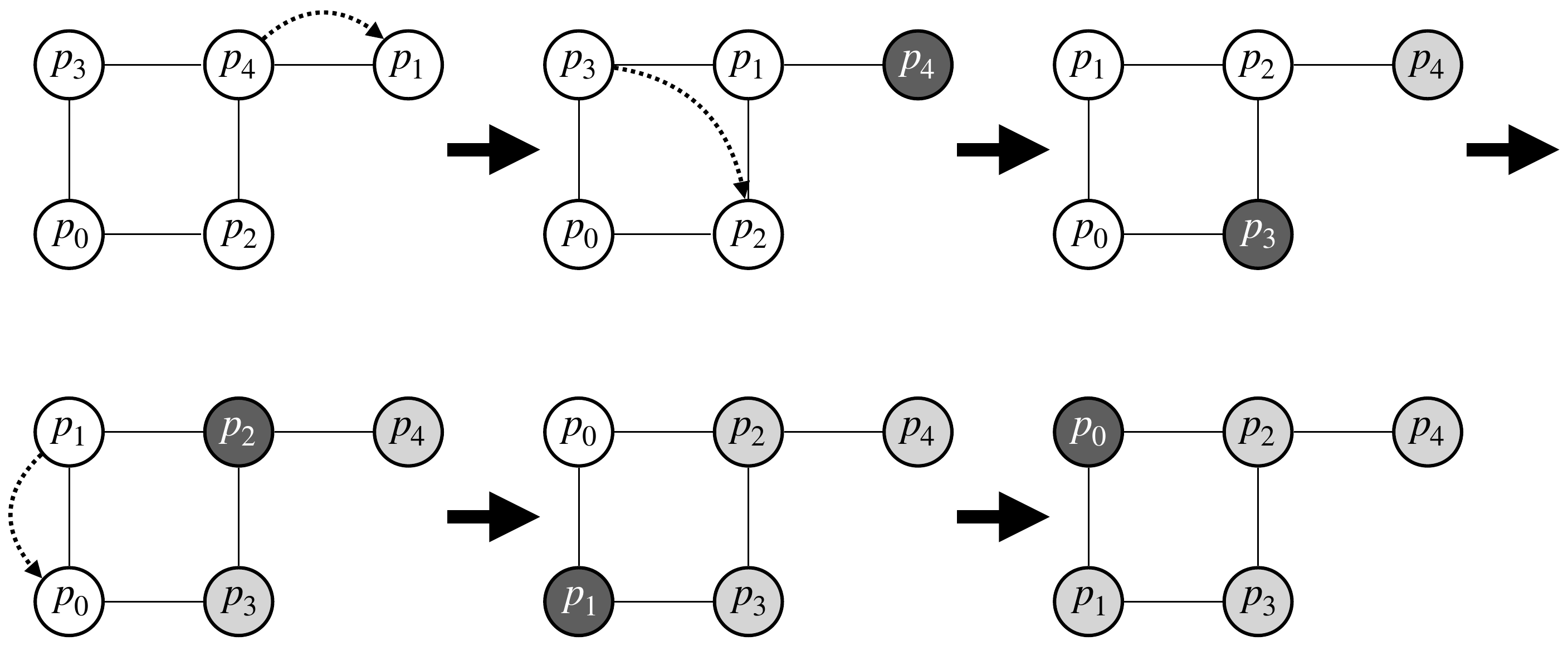}
    \caption{
    Heuristic solution for the token swapping problem. ISAAQ virtually moves physical qubits and fixes them to perform the arrangement. In this example, ISAAQ moves the qubits in the order $p_4, p_3, p_2, p_1, p_0$.
    }
    \label{fig:routing_approx}
\end{figure} 
 
\section{Experimental Results}
\label{section:experimental_results}

\subsection{Implementation Details}
In the implementation of ISAAQ, we use Python 3.10 and Fixstars Amplify Annealing Engine as the backend Ising machine.
Fixstars Amplify Annealing Engine, built using NVIDIA A100, is able to solve QUBO problems that contain hundreds of thousands of binary variables.
To optimize compilation efficiency and quality, we limit the time spent on solving each QUBO problem to 1000 ms and limit the number of logical CNOT gates in each circuit layer to 20.
The size of the circuit chunks is adjusted so that each QUBO problem contains no more than 1200 binary variables.

To evaluate ISAAQ, we use a dataset that has been used in previous research~\cite{zulehner2018efficient, cowtan2019qubit, dury2020qubo}, which contains 158 circuits with 3 to 16 logical qubits and 5 to 224028 logical CNOT gates.
We choose IBM QX5 and IBM QX20 as the physical devices, which contain 16 and 20 physical qubits, respectively.
When the device has more physical qubits than logical qubits, we use a subset of the physical qubits equal in number to the logical qubits.

\subsection{Scheduling Strategies}
We compare the independent, sequential, and binary scheduling strategies in terms of execution time and compilation cost.
To evaluate these strategies, we use IBM QX5 as the physical device and use ``urf2\_277'' from the dataset as the test circuit.
This circuit, which contains 8 logical qubits and 10066 CNOT gates, is divided into 28 circuit chunks.

\begin{figure*}[t]
\begin{minipage}[c]{0.49\linewidth}
        \centering
        \includegraphics[width=0.9\linewidth]{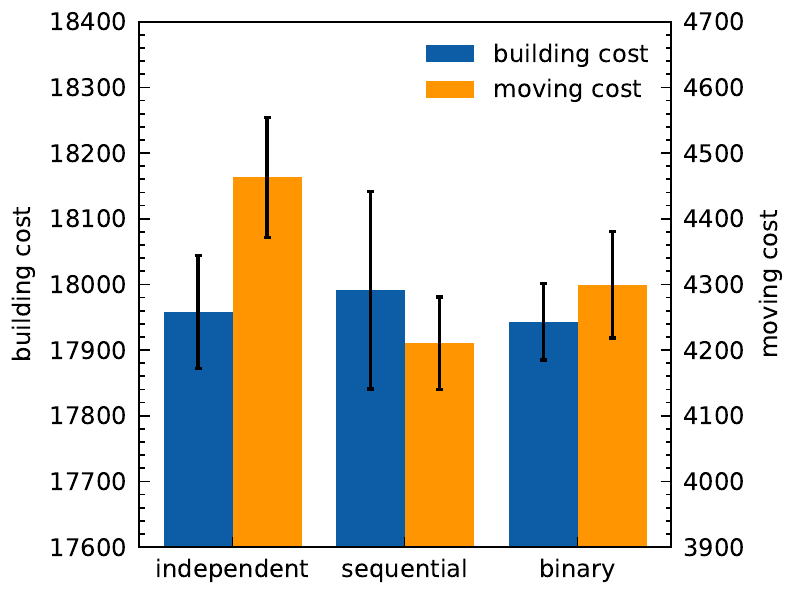}
\subcaption{}
        \label{fig:result_scheduling_cost}
    \end{minipage}
    \begin{minipage}[c]{0.49\linewidth}
        \centering
        \includegraphics[width=0.9\linewidth]{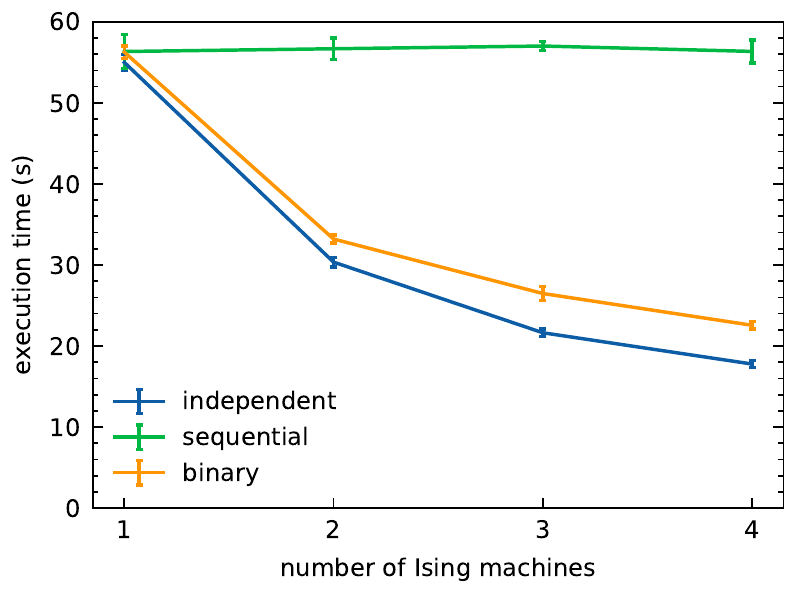}
\subcaption{}
        \label{fig:result_scheduling_time}
    \end{minipage}
    \caption{
    Comparison of scheduling strategies.
    We use ``urf2\_277'' (8 qubits, 10066 CNOT gates) as the original circuit and ``IBM QX5'' as the destination.
    We conduct 5 runs of ISAAQ for each scheduling strategy and each number of Ising machines.
    (a) Comparison based on compilation costs. Both the sequential and binary strategies reduce moving costs, whereas the binary strategy suppresses the variance of building costs. (b) Comparison based on execution time. Both the independent and binary strategies reduce the execution time when using multiple Ising machines.
    }
    \label{fig:result_scheduling}
\end{figure*}

The results for the compilation cost are presented in Fig.~\ref{fig:result_scheduling_cost}.
Both the sequential and binary strategies reduce the moving costs, with the binary strategy also demonstrating a lower variance in the building cost.
Figure~\ref{fig:result_scheduling_time} illustrates the effect of parallelization on execution time when using multiple Ising machines.
Both the independent and binary strategies reduce the execution time as the number of Ising machines increases.
However, the sequential strategy is not able to parallelize qubit routing, even when multiple Ising machines are available.

Overall, these experimental results are consistent with the theoretical results presented in Section \ref{section:isaaq}.
Based on these findings, we conclude that the binary strategy achieves both parallelization and cost reduction among the three scheduling strategies.

\subsection{Moving Cost Approximation}
We evaluate the impact of updating the QUBO model on the accuracy of the approximation of the moving costs.
We use IBM QX5 as the physical device and ``hwb6\_56'' from the dataset as the test circuit. This circuit has 7 logical qubits and 2952 logical CNOT gates.
Because the number of qubits is small enough, we can use the exact solution to calculate the minimum number of required SWAP gates for all possible permutations.
These calculation results are used to create an initial QUBO model.
We then compile the circuit 20 times and update the QUBO model after each compilation, obtaining a sequence of 21 QUBO models.

\begin{figure}[t]
    \centering
    \includegraphics[width=\linewidth]{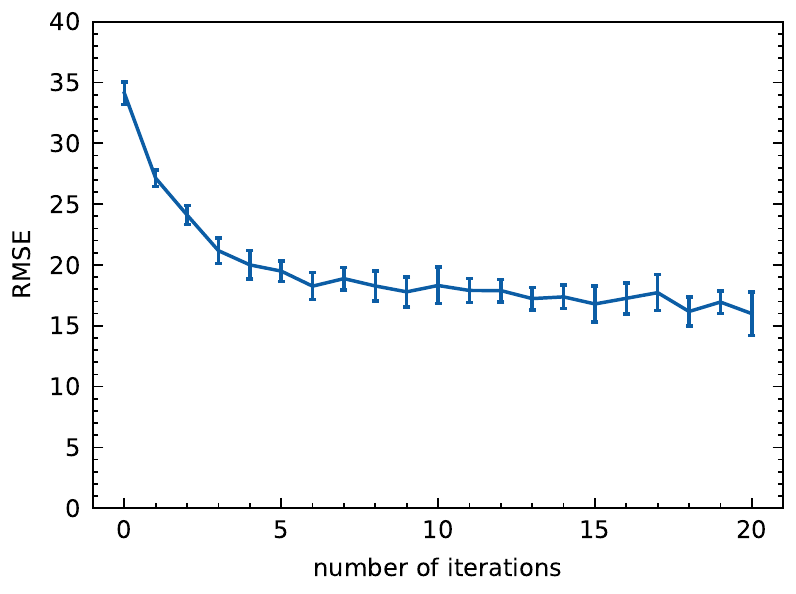}
    \caption{
    Change in Root Mean Squared Error (RMSE) of the approximation of moving costs.
    We use ``hwb6\_56'' (7 qubits, 2952 CNOT gates) as the original circuit and ``IBM QX5'' as the destination.
    We conduct 10 runs of ISAAQ for each QUBO model.
    This result confirms that the incorporation of previous compilation results improves the accuracy of the approximation.
    }
    \label{fig:result_swap_estimation}
\end{figure}

To compare these models, we compile the circuit 10 times using each model and calculate the Root Mean Squared Error (RMSE) of the approximation of the moving costs, defined as
\begin{equation}
\mathrm{RMSE} = \sqrt{\frac{1}{\|S\|}\sum_{\bm{\pi} \in S} \left( 3\mathrm{N_s}(\bm{\pi}) - 3\sum_{\mu=0}^{N-1} a_{\mu,\pi_\mu} \right)^2},
\end{equation}
where $S$ denotes the set of rearrangements obtained from the compilation results.
As shown in Fig.~\ref{fig:result_swap_estimation}, the RMSE of the moving cost approximation decreases with the accumulation of compilation results, indicating that the accuracy of the approximation improves as more data are collected.
Furthermore, this result suggests that the sufficient quantity of compilation result data for tuning the QUBO model constitutes approximately 30\% of the initial data, as the RMSE does not change significantly after 10 iterations.

\begin{figure*}[t]
\begin{minipage}[c]{0.49\linewidth}
        \centering
        \includegraphics[width=0.9\linewidth]{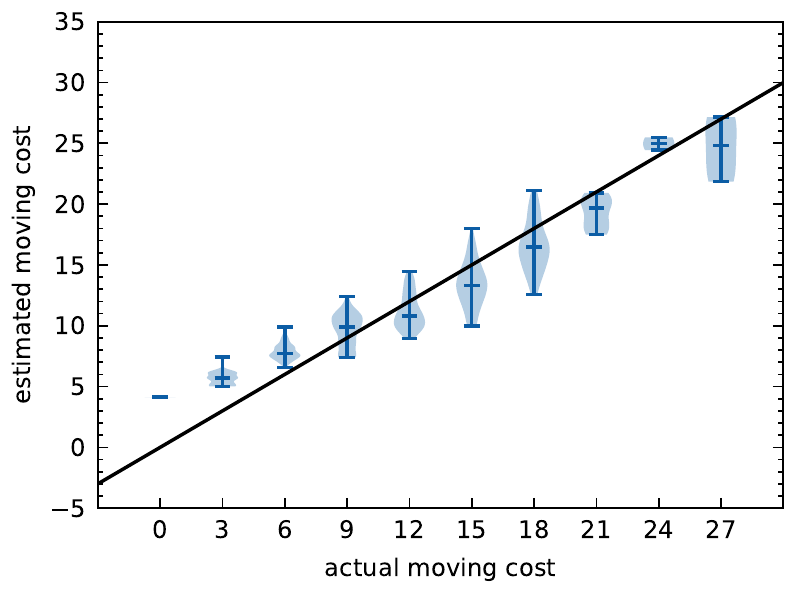}
\subcaption{}
    \end{minipage}
    \begin{minipage}[c]{0.49\linewidth}
        \centering
        \includegraphics[width=0.9\linewidth]{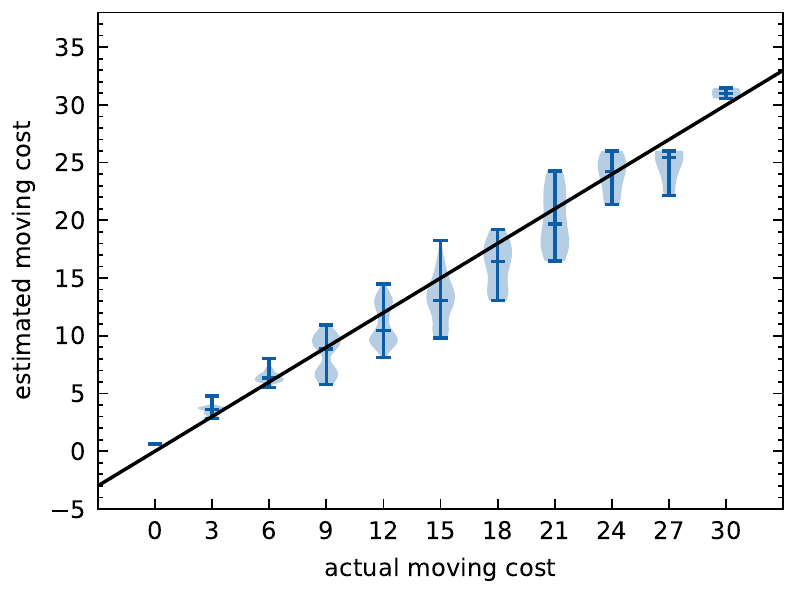}
\subcaption{}
    \end{minipage}
    \caption{
    Comparison of the distributions of estimated moving costs between circuit layers. Black lines represent the line $y = x$ in both figures.
    We use ``hwb6\_56'' (7 qubits, 2952 CNOT gates) as the original circuit and ``IBM QX5'' as the destination.
    We collect the actual moving costs from 10 runs of ISAAQ for each QUBO model.
    (a) Estimated moving costs produced by the initial QUBO model. The initial QUBO model overestimates the moving costs, particularly when they were low. (b) Estimated moving costs produced by the latest QUBO model. The latest QUBO model accurately estimates the moving costs even when they were low.
    }
    \label{fig:result_learning_stitching}
\end{figure*}

We also compare the distributions of the moving costs between circuit layers that are estimated by the initial and latest QUBO models.
Figure~\ref{fig:result_learning_stitching} displays the result, where the x-axis and y-axis represent the minimum moving cost and the estimated moving cost, respectively.
One interesting observation from the result is that the distribution of rearrangements is not uniform, with most rearrangements obtained from ISAAQ requiring a small number of SWAP gates.
The result shows that the latest QUBO models are particularly superior for these rearrangements.
This suggests that the addition of a sufficient quantity of compilation results improves the accuracy of the approximation of the moving costs by correcting the distribution of rearrangements.

\subsection{Building Cost Reduction Method}
We evaluate the effectiveness of the cost-reduction method used in the circuit layer construction step by comparing it with the straightforward method.
To determine which physical device is suitable for the cost-reduction method, we compile ``hwb6\_56'' 10 times on three different devices: Linear, IBM QX5, and IBM QX20.

\begin{figure}[t]
\includegraphics[width=\linewidth]{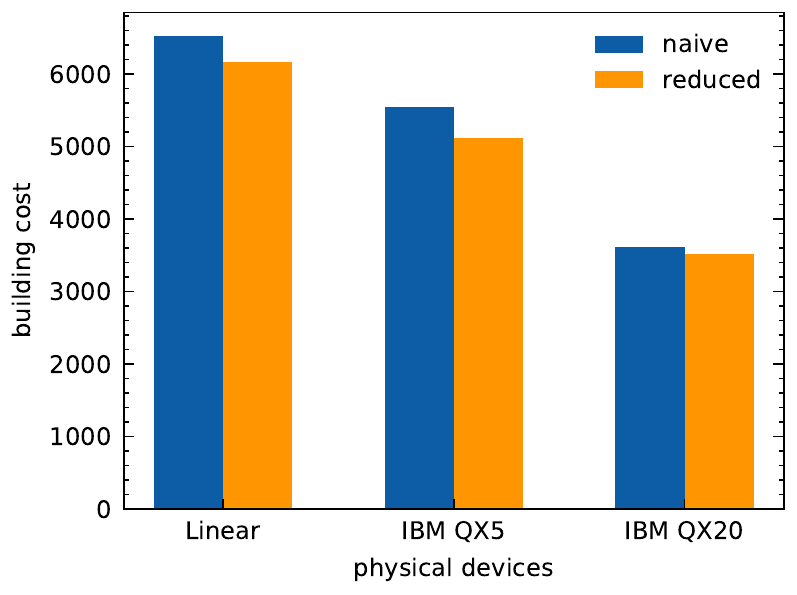}
    \caption{
    Building cost of the compilation of ``hwb6\_56'' to various physical devices.
    We conduct 10 runs of ISAAQ to compile ``hwb6\_56'' (7 qubits, 2952 CNOT gates) on these devices.
    The cost-reduction method reduces the number of additional CNOT gates by 10.1\%, 16.4\%, and 15.8\% for each device, respectively.
    These results reflect the difference in the topology of physical devices.
    }
    \label{fig:result_cost_reduction}
\end{figure}

The results, shown in Fig.~\ref{fig:result_cost_reduction}, indicate that both methods require the largest number of physical CNOT gates for the Linear device and the smallest number for the IBM QX20 device.
This occurrence suggests that the average distance between physical qubits affects compilation quality, as the physical qubits on the Linear device are farther apart whereas those on the IBM QX20 device are densely connected.

The results also show that the cost-reduction method reduces the number of additional CNOT gates by 10.1\%, 16.4\%, and 15.8\% for each device, respectively.
This outcome suggests that the method is particularly effective for planar devices, as it functions well when many logical qubits are placed near the relay qubit.
However, on the Linear device, logical qubits cannot be placed close to each other because each physical qubit only has at most two neighbors.

\subsection{Comparison with Other Compilers}
To evaluate the effectiveness of ISAAQ, we use it to compile the 158 circuits in the dataset for both IBM QX5 and IBM QX20.
We then compare the performance of ISAAQ to that of previous heuristic methods presented by Li et al.~\cite{li2019tackling} and Cowtan et al.~\cite{cowtan2019qubit}, as well as the QUBO method presented by Dury et al.~\cite{dury2020qubo}.
We measure the performance of each compiler based on the \textit{average compilation cost}, which is defined as the ratio of the number of used physical CNOT gates to the number of logical CNOT gates. 
To measure the average compilation costs for Li et al.'s method, we compile the circuit with their method implemented in Qiskit.
For the other methods, we use the experimental data listed in the respective papers.
Due to runtime errors in Qiskit and the lack of compilation results, 21 circuits are excluded from the evaluation for IBM QX5 and 14 circuits are excluded for IBM QX20.
To investigate for which circuits the advantages of ISAAQ are specific, the circuits are classified into three categories based on the number of logical CNOT gates: small (less than 100), intermediate (between 100 and 1000), and large (more than 1000).

\begin{figure*}[t]
\begin{minipage}[c]{0.32\linewidth}
        \centering
        \includegraphics[width=0.9\linewidth]{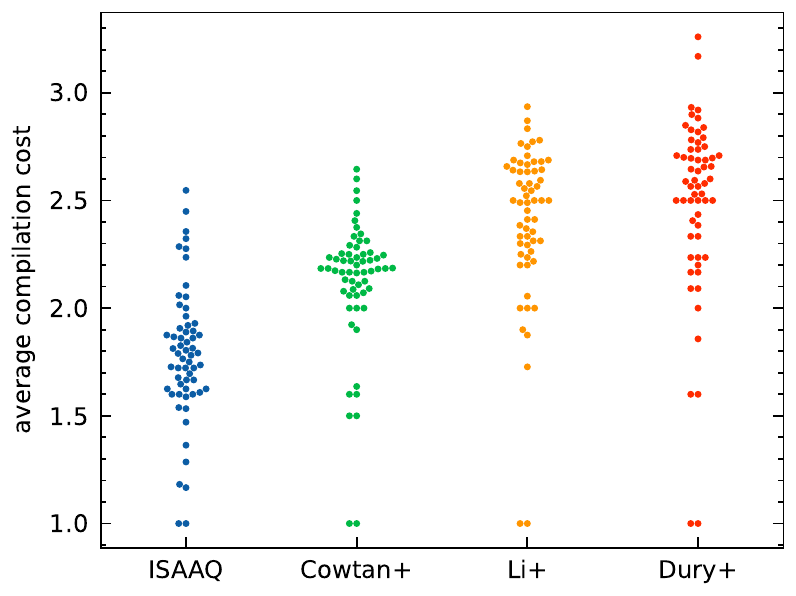}
\subcaption{}
    \end{minipage}
    \begin{minipage}[c]{0.32\linewidth}
        \centering
        \includegraphics[width=0.9\linewidth]{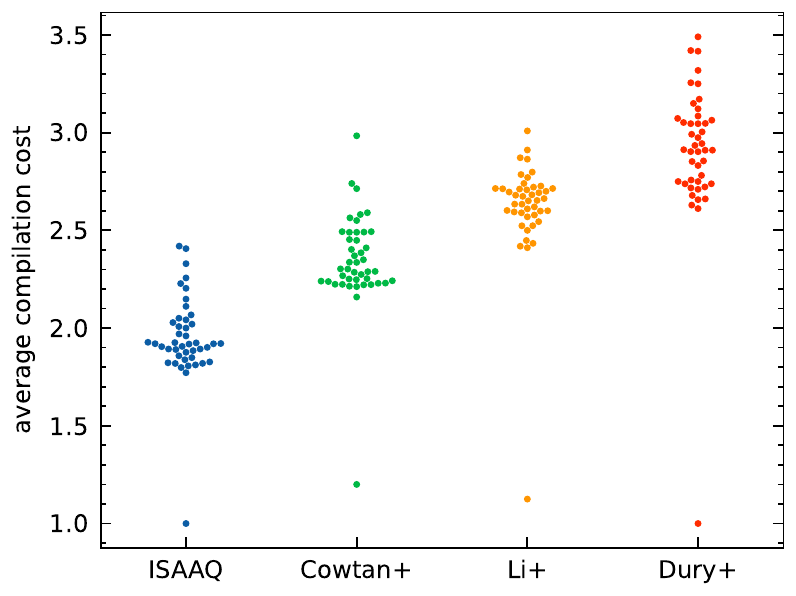}
\subcaption{}
    \end{minipage}
    \begin{minipage}[c]{0.32\linewidth}
        \centering
        \includegraphics[width=0.9\linewidth]{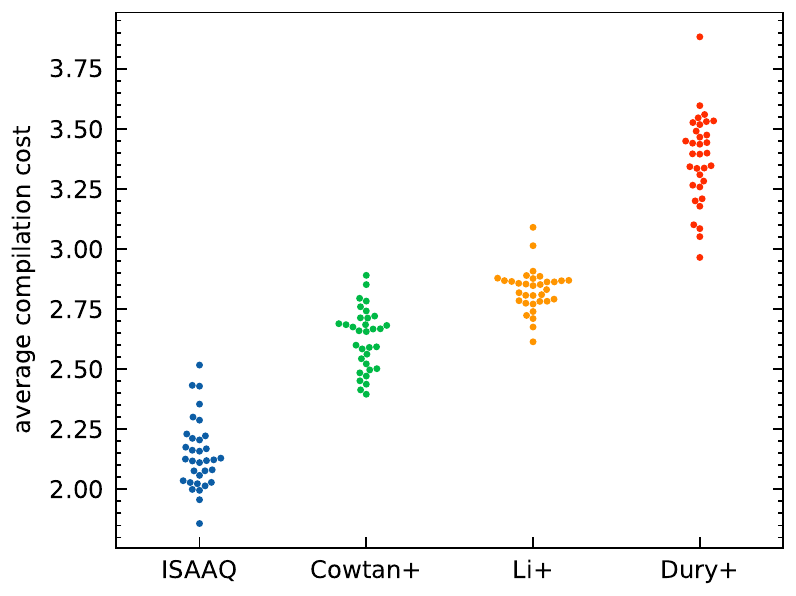}
\subcaption{}
    \end{minipage}
    \caption{
    Comparison of ISAAQ with other compilers based on the distribution of average compilation costs for IBM QX5.
    The circuits are categorized into (a) ``small'' (60 circuits), (b) ``intermediate'' (44 circuits), and (c) ``large'' (33 circuits).
    }
    \label{fig:result_melbourne}
\end{figure*}

\begin{figure*}[t]
\begin{minipage}[c]{0.32\linewidth}
        \centering
        \includegraphics[width=0.9\linewidth]{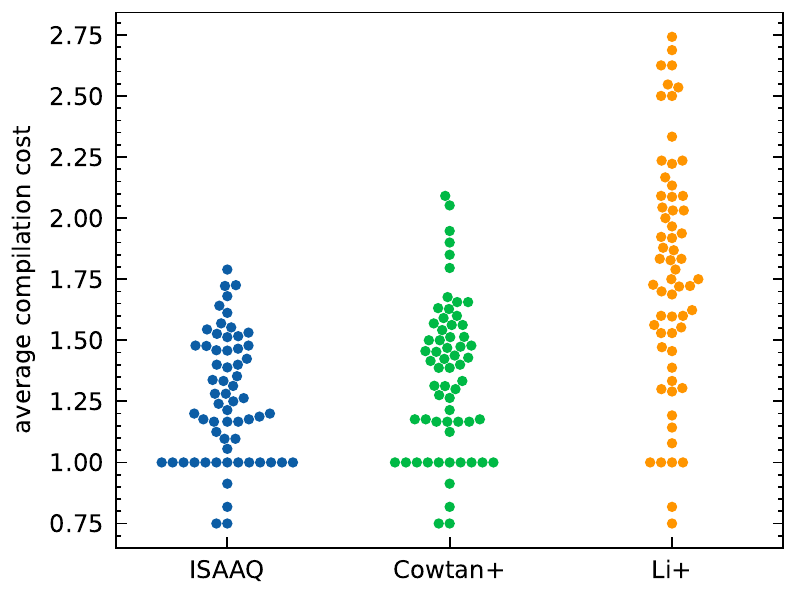}
\subcaption{}
    \end{minipage}
    \begin{minipage}[c]{0.32\linewidth}
        \centering
        \includegraphics[width=0.9\linewidth]{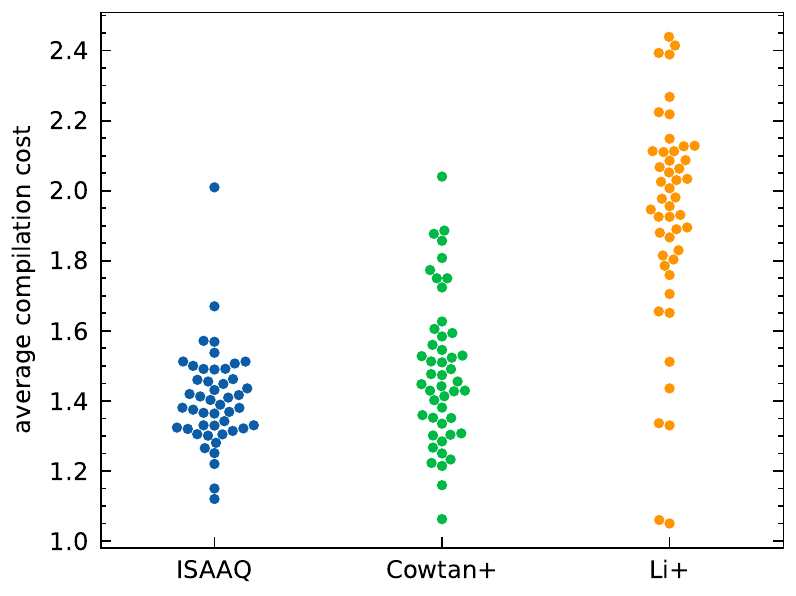}
\subcaption{}
    \end{minipage}
    \begin{minipage}[c]{0.32\linewidth}
        \centering
        \includegraphics[width=0.9\linewidth]{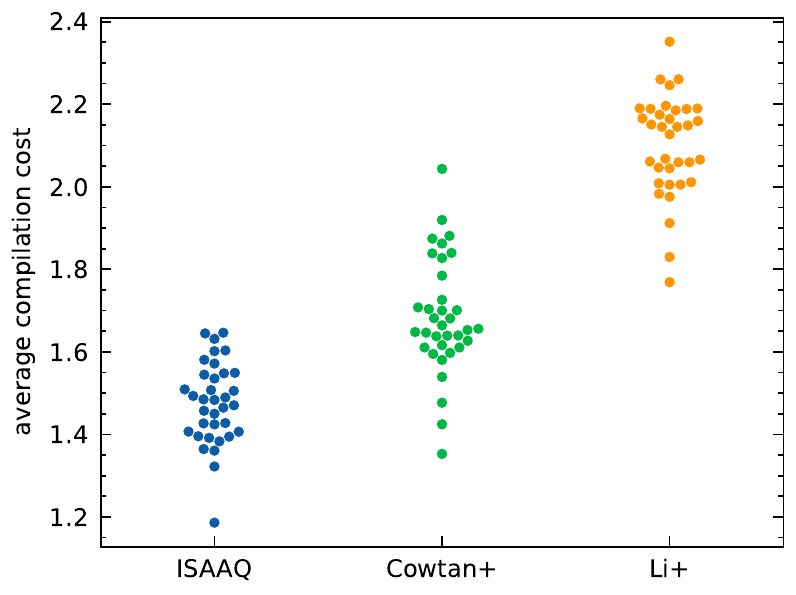}
\subcaption{}
    \end{minipage}
    \caption{
    Comparison of ISAAQ with other compilers based on the distribution of average compilation costs for IBM QX20.
    The circuits are categorized into (a) ``small'' (62 circuits), (b) ``intermediate'' (47 circuits), and (c) ``large'' (35 circuits).
    }
    \label{fig:result_tokyo}
\end{figure*}

As shown in Fig.~\ref{fig:result_melbourne} and Fig.~\ref{fig:result_tokyo}, the distributions of the average compilation costs indicate that ISAAQ consistently performs the most efficient compilation for all categories of circuits on both IBM QX5 and IBM QX20.
The superior performance of ISAAQ is particularly evident on IBM QX5, where ISAAQ achieves average compilation costs below 2.0 for more than half of the circuits, whereas the other methods incur average compilation costs above 2.0 for almost all circuits.
These results also show that the difference in compilation efficiency becomes more apparent as the size of the circuit increases, indicating that ISAAQ is particularly effective for large circuits.

\begin{figure*}[t]
\begin{minipage}[c]{0.49\linewidth}
        \centering
        \includegraphics[width=0.9\linewidth]{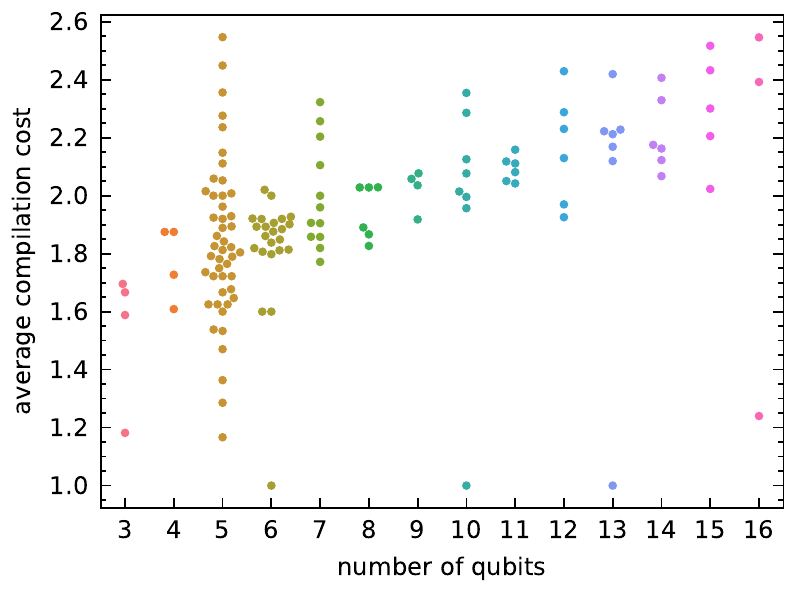}
\subcaption{}
    \end{minipage}
    \begin{minipage}[c]{0.49\linewidth}
        \centering
        \includegraphics[width=0.9\linewidth]{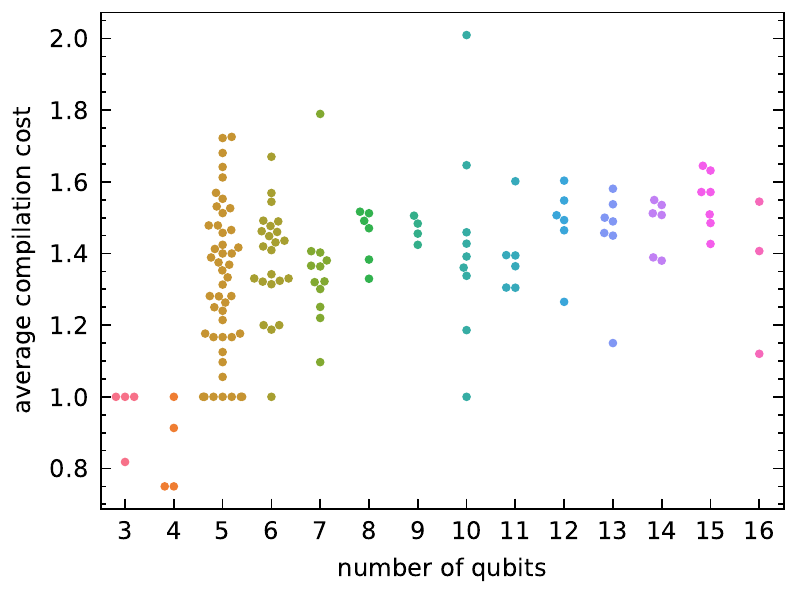}
\subcaption{}
    \end{minipage}
    \caption{
    Average compilation cost with respect to the number of qubits for (a) IBM QX5 and (b) IBM QX20.
    The increase in average compilation cost is almost constant for IBM QX5 and shows a slowdown for IBM QX20.
    }
    \label{fig:result_qubit}
\end{figure*}

\begin{figure*}[t]
\begin{minipage}[c]{0.49\linewidth}
        \centering
        \includegraphics[width=0.9\linewidth]{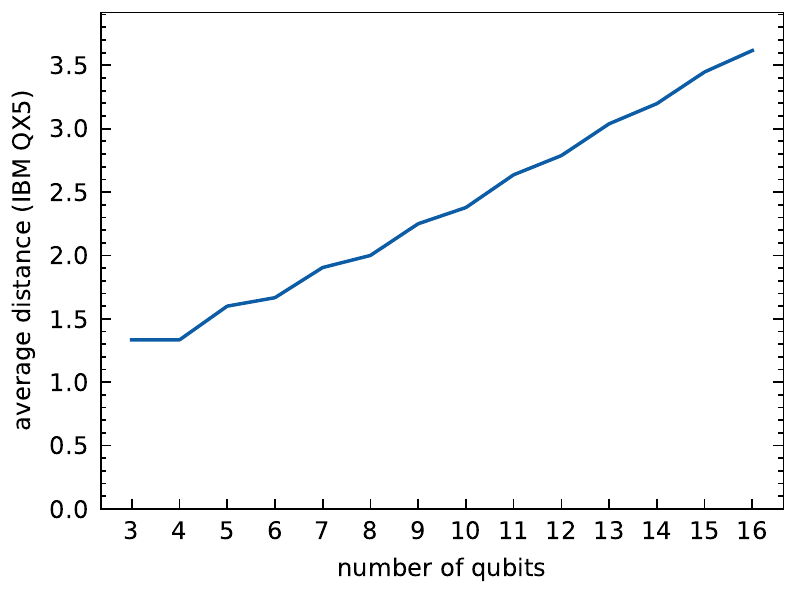}
\subcaption{}
    \end{minipage}
    \begin{minipage}[c]{0.49\linewidth}
        \centering
        \includegraphics[width=0.9\linewidth]{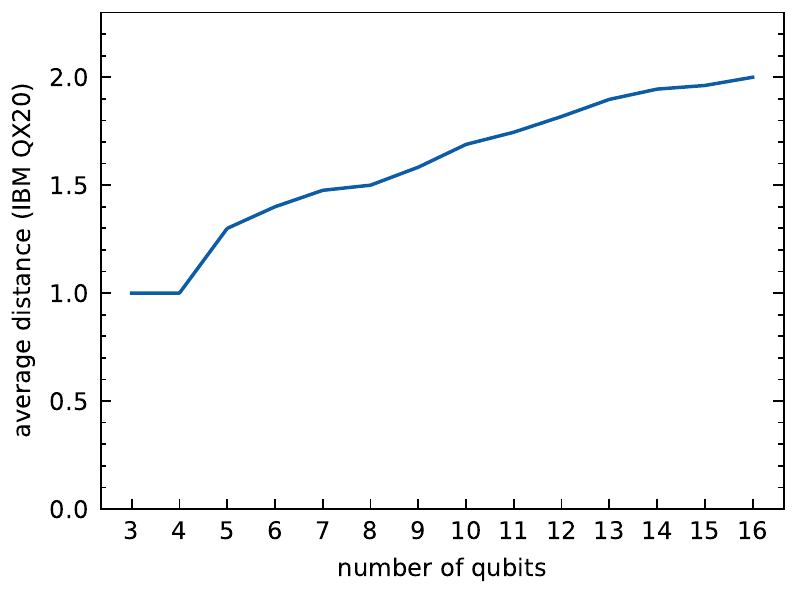}
\subcaption{}
    \end{minipage}
    \caption{
    Average distance between physical qubits in (a) IBM QX5 and (b) IBM QX20.
    Due to the difference in the connectivity of physical qubits, the average distance continues increasing at a constant rate for IBM QX5 whereas the rate of increase levels off slowing down for IBM QX20.
    }
    \label{fig:result_average_distance}
\end{figure*}

The results suggest that the average compilation costs are affected by the average distance between the physical qubits.
A comparison of Fig.~\ref{fig:result_melbourne} and Fig.~\ref{fig:result_tokyo} indicates that the average compilation costs for IBM QX20 are lower than for IBM QX5 for all categories, which indicates that the difference in the average distance between physical qubits had a significant impact on compilation efficiency.
Figure~\ref{fig:result_qubit} illustrates the relationship between the average compilation costs and the number of qubits, showing that the increase in the average compilation cost remains almost constant for IBM QX5, whereas it slows down for IBM QX20.
This trend is consistent with that of the average distance between physical qubits shown in Fig.~\ref{fig:result_average_distance}, indicating that these two factors are correlated.

\section{Discussion}
\label{section:discussions}
The results of our study demonstrate that ISAAQ outperforms other compilers in terms of efficiency in quantum circuit compilation and scalability with respect to the number of logical CNOT gates.
The superior performance of ISAAQ can be attributed to its accurate approximation of the moving costs, efficient implementation of logical CNOT gates, and ability to perform global optimization in qubit routing.
ISAAQ adjusts the distribution of rearrangements of logical qubits after compilation, which indicates that ISAAQ requires a measure of compilation experience to reach its maximum performance.
The cost-reducing effect of the proposed implementation method is demonstrated to be more significant for IBM QX5 and IBM QX20, suggesting that the use of relay qubits can improve compilation quality for planar devices.
Unlike the routing methods proposed by Li et al. and Cowtan et al., which can only optimize the placement of logical qubits locally, ISAAQ uses Ising machines to solve QUBO problems that represent qubit routing problems over large parts of the circuit, resulting in a global optimal solution.
The scalability of ISAAQ derives from the ability of Ising machines to efficiently find solutions to large QUBO problems, which suggests that ISAAQ can be improved with further progress in the capabilities of Ising machines.

Despite its advantages over other compilers, ISAAQ faces a challenge with respect to scaling the number of qubits.
As the number of qubits increases, the number of binary variables used in a single QUBO problem increases quadratically.
To address this issue, it is crucial to optimize the QUBO formulation to express the qubit routing problem using a smaller number of binary variables.
Another area of potential improvement is the estimation of the building cost. 
While ISAAQ can accurately estimate the moving costs by updating QUBO models using previous compilation results, it remains reliant on the straightforward method to estimate the building cost, neglecting to take the cost-reduction techniques used in the circuit construction step into account.
An accurate estimation of the compilation cost could enable Ising machines to find better solutions, leading to better compilation quality.
Furthermore, ISAAQ only uses the basic characteristics of CNOT gates for optimization and ignores other properties including commutativity and composition rules.
As shown in previous studies, the utilization of such properties can simplify circuits and improve overall performance~\cite{itoko2020optimization, maslov2008quantum}. 

\section{Conclusions}
\label{section:conclusion}
We introduce ISAAQ, a quantum compiler that leverages the power of Ising machines to reduce the number of physical CNOT gates needed for qubit routing.
By representing the routes of logical qubits as a sequence of mappings, we formulate the compilation cost as the sum of the building and moving costs, which can be expressed in QUBO form.
ISAAQ performs qubit routing with multiple Ising machines and updates itself after compiling circuits for maximum accuracy, which contributes to the effective utilization of Ising machines.
We propose a cost-reduction method for the efficient synthesis of logical CNOT gates, which is demonstrated to work effectively for planar devices.
Our results show that ISAAQ outperforms existing heuristic and QUBO methods for most circuits, on both IBM QX5 and IBM QX20.
This superior performance is particularly noticeable for large circuits, indicating ISAAQ's ability to achieve global optimization by using Ising machines.

ISAAQ illustrates the potential of Ising machines as a key tool in quantum compilers, as they can efficiently solve various combinatorial optimization problems, including qubit routing.
The results of this study are expected to provide insight into the development of more efficient and powerful quantum compilers, which are essential for the future era of quantum computing.

\begin{acknowledgements}
S. T. acknowledges support by JSPS KAKENHI Grant Number JP21K03391, JST, CREST Grant Number JPMJCR19K4, Japan, and JST Grant Number JPMJPF2221.
Y. H. acknowledges support by JSPS KAKENHI Grant Number JP22H03659.
\end{acknowledgements}
 
\appendix
\section{Optimal QUBO Coefficients for Moving Cost Approximation}
\label{section:determining_qubo_coefficients}
In determining the coefficients $a_{\mu, \nu}$ in Eq.~\eqref{eq:SWAP_cost_approx}, we aim to maximize the accuracy of the approximation.
The squared error $E$ is defined as
\begin{align}
E &\coloneqq \frac{1}{2} \sum_{\bm{\pi} \in S} \left( \hat{\mathrm{N_s}}(\bm{\pi}) - \sum_{\mu=0}^{N-1} a_{\mu,\pi_\mu} \right)^2,
\end{align}
where $S$ and $\hat{\mathrm{N_s}}(\bm{\pi})$ represent the set of logical qubit rearrangements obtained from ISAAQ and the number of SWAP gates used to implement $\bm{\pi}$, respectively.
Because $E$ is convex with respect to $a_{\mu, \nu}$, the optimal coefficients can be obtained by solving
\begin{align}
\frac{\partial E}{\partial a_{\mu,\nu}} = \sum_{\substack{\bm{\pi} \in S \\ \pi_\mu = \nu}} \left(\sum_{\mu'=0}^{N-1} a_{\mu',\pi_{\mu'}} - \hat{\mathrm{N_s}}(\bm{\pi}) \right) = 0
\end{align}
for every $\mu$ and $\nu$.
As this equation degenerates, we solve the following constrained optimization problem:
\begin{align}
\mathrm{minimize} &\quad \sum_{\mu=0}^{N-1}\sum_{\nu=0}^{N-1}a_{\mu,\nu}^2 \nonumber \\
\mathrm{subject\;to} &\quad \sum_{\mu'=0}^{N-1} \sum_{\nu'=0}^{N-1} \sum_{\substack{\bm{\pi} \in S \\ \pi_\mu = \nu \\ \pi_{\mu'} = \nu'}} a_{\mu',\nu'}  = \sum_{\substack{\bm{\pi} \in S \\ \pi_\mu = \nu}} \hat{\mathrm{N_s}}(\bm{\pi})
\label{eq:linear_equations}
\end{align}
to obtain the solution that minimizes the squared norm $\|\bm{a}\|^2 = \sum_{\mu=0}^{N-1}\sum_{\nu=0}^{N-1}a_{\mu,\nu}^2$.

In the absence of previous compilation results, we assume that $\bm{\pi}$ follows a uniform distribution of permutations and use the set of all permutations $\Pi$ instead of $S$.
Using this assumption, we can transform Eq.~\eqref{eq:linear_equations} as follows:
\begin{align}
\sum_{\mu'=0}^{N-1} \sum_{\nu'=0}^{N-1} \sum_{\substack{\bm{\pi} \in S \\ \pi_\mu = \nu \\ \pi_{\mu'} = \nu'}} a_{\mu',\nu'}
&=
(N-1)! a_{\mu,\nu} + \sum_{\substack{\mu' \neq \mu \\ \nu' \neq \nu}} (N-2)! a_{\mu',\nu'}, \nonumber \\
\sum_{\substack{\bm{\pi} \in S \\ \pi_\mu = \nu}} \hat{\mathrm{N_s}}(\bm{\pi})
&=
(N-1)! \langle \hat{\mathrm{N_s}}(\bm{\pi}) \rangle_{\pi_\mu = \nu},
\end{align}
which leads to
\begin{align}
a_{\mu,\nu} + \frac{1}{N-1} \sum_{\substack{\mu' \neq \mu \\ \nu' \neq \nu}} a_{\mu',\nu'} = \langle \hat{\mathrm{N_s}}(\bm{\pi}) \rangle_{\pi_\mu = \nu}
\label{eq:uniform_linear_equations}
\end{align}
for every $\mu$ and $\nu$.
Summing over $a_{\mu, \nu}$ with respect to $\mu$ and $\nu$ provides the following constraint:
\begin{align}
\sum_{\mu, \nu} a_{\mu, \nu} = N \langle \hat{\mathrm{N_s}}(\bm{\pi}) \rangle.
\end{align}

To investigate Eq.~\eqref{eq:uniform_linear_equations}, we use variables $R_\mu \coloneqq \sum_{\nu'} a_{\mu, \nu'}$ and $C_\nu \coloneqq \sum_{\mu'} a_{\mu', \nu}$, which represent the sum of each row and column.
Eq.~\eqref{eq:uniform_linear_equations} can be transformed as follows:
\begin{align}
a_{\mu,\nu} = \frac{N-1}{N} \langle \hat{\mathrm{N_s}}(\bm{\pi}) \rangle_{\pi_\mu = \nu} - \langle \hat{\mathrm{N_s}}(\bm{\pi}) \rangle + \frac{1}{N} \left( R_\mu + C_\nu \right),
\label{eq:flat_explicit_form}
\end{align}
which explicitly represents $a_{\mu, \nu}$ using $R_\mu$ and $C_\nu$.
This equation implies that
\begin{align}
\sum_{\mu'} a_{\mu',\nu} &= C_\nu - \langle \hat{\mathrm{N_s}}(\bm{\pi}) \rangle + \frac{1}{N} \sum_{\mu'} R_{\mu'}, \nonumber \\
\sum_{\nu'} a_{\mu,\nu'} &= R_\mu - \langle \hat{\mathrm{N_s}}(\bm{\pi}) \rangle + \frac{1}{N} \sum_{\nu'} C_{\nu'},
\end{align}
which indicates that the condition
\begin{align}
\frac{1}{N} \sum_{\mu'} R_{\mu'} = \frac{1}{N} \sum_{\nu'} C_{\nu'} = \langle \hat{\mathrm{N_s}}(\bm{\pi}) \rangle
\end{align}
is required for consistency.
This condition is necessary and sufficient for generating any solution for Eq.~\eqref{eq:uniform_linear_equations}.

We solve the constrained optimization problem in Eq.~\eqref{eq:linear_equations} using the method of Lagrange multipliers.
The Lagrangian function is given as
\begin{align}
L &= \|\bm{a}\|^2 - \lambda_r \left(\sum_\mu \frac{R_\mu}{N} - \langle \hat{\mathrm{N_s}}(\bm{\pi}) \rangle \right) - \lambda_c \left( \sum_\mu \frac{C_\nu}{N} - \langle \hat{\mathrm{N_s}}(\bm{\pi}) \rangle \right),
\end{align}
where $\lambda_r$ and $\lambda_c$ are the Lagrange multipliers.
By applying Eq.~\eqref{eq:flat_explicit_form}, the derivatives of the Lagrangian function with respect to $R_\mu$ and $C_\nu$ can be obtained as follows:
\begin{align}
\frac{\partial L}{\partial R_\mu}
&= \frac{2}{N} R_\mu - \frac{\lambda_r}{N}, \nonumber \\
\frac{\partial L}{\partial C_\nu}
&= \frac{2}{N} C_\nu - \frac{\lambda_c}{N}.
\end{align}
Solving $\frac{\partial L}{\partial R_\mu} = \frac{\partial L}{\partial C_\nu} = 0$ provides the following condition for minimizing the Lagrangian function
\begin{align}
R_\mu = \frac{\lambda_r}{2} = \langle \hat{\mathrm{N_s}}(\bm{\pi}) \rangle, \nonumber \\
C_\nu = \frac{\lambda_c}{2} = \langle \hat{\mathrm{N_s}}(\bm{\pi}) \rangle,
\end{align}
which shows that the optimal coefficients for $a_{\mu, \nu}$ can be calculated as
\begin{align}
a_{\mu, \nu} = \frac{N-1}{N} \langle \hat{\mathrm{N_s}}(\bm{\pi}) \rangle_{\pi_\mu = \nu} - \frac{N-2}{N} \langle \hat{\mathrm{N_s}}(\bm{\pi}) \rangle
\end{align}
without using any previous compilation results.

\end{document}